\documentclass[aps,prc,reprint,groupedaddress,superscriptaddress]{revtex4-1}
\usepackage{amsmath}
\usepackage{graphicx}
\usepackage{array}
\usepackage{capt-of}
\usepackage{float}
\usepackage{placeins}
\usepackage{longtable}

\begin{document}



\title{Experimental study of the $\gamma p\rightarrow K^0\Sigma^+$, $\gamma n\rightarrow K^0\Lambda$, and $\gamma n\rightarrow K^0 \Sigma^0$ reactions at the Mainz Microtron}



\author{C.S.~Akondi}
\author{K.~Bantawa}
\author{D.M.~Manley}
\address{Kent State University, Kent, Ohio 44242, USA}

\author{S.~Abt}
\address{Department of Physics, University of Basel, Ch-4056 Basel, Switzerland}

\author{P.~Achenbach}
\address{Institut f\"ur Kernphysik, University of Mainz, D-55099 Mainz, Germany}

\author{F.~Afzal}
\address{Helmholtz-Institut f\"ur Strahlen- und Kernphysik, University Bonn, D-53115 Bonn, Germany}

\author{P.~Aguar-Bartolom\'e}
\address{Institut f\"ur Kernphysik, University of Mainz, D-55099 Mainz, Germany}

\author{Z.~Ahmed}
\address{University of Regina, Regina, SK S4S-0A2 Canada}

\author{J.R.M.~Annand}
\address{SUPA School of Physics and Astronomy, University of Glasgow, Glasgow, G12 8QQ, UK}

\author{H.J.~Arends}
\address{Institut f\"ur Kernphysik, University of Mainz, D-55099 Mainz, Germany}

\author{M.~Bashkanov}
\address{Department of Physics, University of York, Heslington, York, YO10 5DD, UK}

\author{R.~Beck}
\address{Helmholtz-Institut f\"ur Strahlen- und Kernphysik, University Bonn, D-53115 Bonn, Germany}

\author{M.~Biroth}
\address{Institut f\"ur Kernphysik, University of Mainz, D-55099 Mainz, Germany}

\author{N.~Borisov}
\address{Joint Institute for Nuclear Research, 141980 Dubna, Russia}

\author{A.~Braghieri}
\address{INFN Sezione di Pavia, I-27100 Pavia, Pavia, Italy}

\author{W.J.~Briscoe}
\address{Center for Nuclear Studies, The George Washington University, Washington, DC 20052, USA}

\author{F.~Cividini}
\address{Institut f\"ur Kernphysik, University of Mainz, D-55099 Mainz, Germany}

\author{C.~Collicott}
\address{Department of Astronomy and Physics, Saint Mary's University, E4L1E6 Halifax, Canada}

\author{S.~Costanza}
\address{Dipartimento di Fisica, Universit\`a di Pavia, I-27100 Pavia, Italy}
\address{INFN Sezione di Pavia, I-27100 Pavia, Pavia, Italy}

\author{A.~Denig}
\address{Institut f\"ur Kernphysik, University of Mainz, D-55099 Mainz, Germany}

\author{M.~Dieterle}
\address{Department of Physics, University of Basel, Ch-4056 Basel, Switzerland}

\author{E.J.~Downie}
\address{Center for Nuclear Studies, The George Washington University, Washington, DC 20052, USA}

\author{P.~Drexler}
\address{Institut f\"ur Kernphysik, University of Mainz, D-55099 Mainz, Germany}

\author{M.I.~Ferretti Bondy}
\address{Institut f\"ur Kernphysik, University of Mainz, D-55099 Mainz, Germany}

\author{S.~Gardner}
\address{SUPA School of Physics and Astronomy, University of Glasgow, Glasgow, G12 8QQ, UK}

\author{D.I.~Glazier}
\address{SUPA School of Physics and Astronomy, University of Glasgow, Glasgow, G12 8QQ, UK}

\author{I.~Gorodnov}
\address{Joint Institute for Nuclear Research, 141980 Dubna, Russia}

\author{W.~Gradl}
\address{Institut f\"ur Kernphysik, University of Mainz, D-55099 Mainz, Germany}

\author{M.~G\"unther}
\address{Department of Physics, University of Basel, Ch-4056 Basel, Switzerland}

\author{D.~Gurevich}
\address{Institute for Nuclear Research, RU-125047 Moscow, Russia}

\author{S.~Kay}
\address{University of Regina, Regina, SK S4S-0A2 Canada}

\author{L.~Heijkenskj{\"o}ld}
\address{Institut f\"ur Kernphysik, University of Mainz, D-55099 Mainz, Germany}

\author{D.~Hornidge}
\address{Mount Allison University, Sackville, New Brunswick E4L1E6, Canada}

\author{G.M.~Huber}
\address{University of Regina, Regina, SK S4S-0A2 Canada}

\author{M.~K\"aser}
\address{Department of Physics, University of Basel, Ch-4056 Basel, Switzerland}

\author{V.L.~Kashevarov}
\address{Institut f\"ur Kernphysik, University of Mainz, D-55099 Mainz, Germany}

\author{M.~Korolija}
\address{Rudjer Boskovic Institute, HR-10000 Zagreb, Croatia}

\author{B.~Krusche}
\address{Department of Physics, University of Basel, Ch-4056 Basel, Switzerland}

\author{A.~Lazarev}
\address{Joint Institute for Nuclear Research, 141980 Dubna, Russia}

\author{K.~Livingston}
\address{SUPA School of Physics and Astronomy, University of Glasgow, Glasgow, G12 8QQ, UK}

\author{S.~Lutterer}
\address{Department of Physics, University of Basel, Ch-4056 Basel, Switzerland}

\author{J.C.~McGeorge}
\address{SUPA School of Physics and Astronomy, University of Glasgow, Glasgow, G12 8QQ, UK}

\author{I.J.D.~MacGregor}
\address{SUPA School of Physics and Astronomy, University of Glasgow, Glasgow, G12 8QQ, UK}

\author{P.P.~Martel}
\address{Institut f\"ur Kernphysik, University of Mainz, D-55099 Mainz, Germany}

\author{D.G.~Middleton}
\address{Institut f\"ur Kernphysik, University of Mainz, D-55099 Mainz, Germany}
\address{Mount Allison University, Sackville, New Brunswick E4L1E6, Canada}

\author{R.~Miskimen}
\address{University of Massachusetts, Amherst, Massachusetts 01003, USA}

\author{E.~Mornacchi}
\address{Institut f\"ur Kernphysik, University of Mainz, D-55099 Mainz, Germany}

\author{A.~Mushkarenkov}
\address{INFN Sezione di Pavia, I-27100 Pavia, Pavia, Italy}
\address{University of Massachusetts, Amherst, Massachusetts 01003, USA}

\author{C. Mullen}
\address{SUPA School of Physics and Astronomy, University of Glasgow, Glasgow, G12 8QQ, UK}

\author{A.~Neganov}
\address{Joint Institute for Nuclear Research, 141980 Dubna, Russia}

\author{A.~Neiser}
\address{Institut f\"ur Kernphysik, University of Mainz, D-55099 Mainz, Germany}

\author{M.~Oberle}
\address{Department of Physics, University of Basel, Ch-4056 Basel, Switzerland}

\author{M.~Ostrick}
\address{Institut f\"ur Kernphysik, University of Mainz, D-55099 Mainz, Germany}

\author{P.B.~Otte}
\address{Institut f\"ur Kernphysik, University of Mainz, D-55099 Mainz, Germany}

\author{B.~Oussena}
\address{Institut f\"ur Kernphysik, University of Mainz, D-55099 Mainz, Germany}
\address{Center for Nuclear Studies, The George Washington University, Washington, DC 20052, USA}

\author{D.~Paudyal}
\address{University of Regina, Regina, SK S4S-0A2 Canada}

\author{P.~Pedroni}
\address{INFN Sezione di Pavia, I-27100 Pavia, Pavia, Italy}

\author{A.~Powell}
\address{SUPA School of Physics and Astronomy, University of Glasgow, Glasgow, G12 8QQ, UK}

\author{S.N.~Prakhov}
\address{Institut f\"ur Kernphysik, University of Mainz, D-55099 Mainz, Germany}
\address{University of California Los Angeles, Los Angeles, California 90095-1547, USA}

\author{G.~Ron}
\address{Racah Institute of Physics, Hebrew University of Jerusalem, Jerusalem 91904, Israel}

\author{T.~Rostomyan}
\address{Department of Physics, University of Basel, Ch-4056 Basel, Switzerland}

\author{A.~Sarty}
\address{Department of Astronomy and Physics, Saint Mary's University, E4L1E6 Halifax, Canada}

\author{C.~Sfienti}
\address{Institut f\"ur Kernphysik, University of Mainz, D-55099 Mainz, Germany}

\author{V.~Sokhoyan}
\address{Institut f\"ur Kernphysik, University of Mainz, D-55099 Mainz, Germany}

\author{K.~Spieker}
\address{Helmholtz-Institut f\"ur Strahlen- und Kernphysik, University Bonn, D-53115 Bonn, Germany}

\author{O.~Steffen}
\address{Institut f\"ur Kernphysik, University of Mainz, D-55099 Mainz, Germany}

\author{I.I.~Strakovsky}
\address{Center for Nuclear Studies, The George Washington University, Washington, DC 20052, USA}

\author{Th.~Strub}
\address{Department of Physics, University of Basel, Ch-4056 Basel, Switzerland}

\author{I.~Supek}
\address{Rudjer Boskovic Institute, HR-10000 Zagreb, Croatia}

\author{A.~Thiel}
\address{Helmholtz-Institut f\"ur Strahlen- und Kernphysik, University Bonn, D-53115 Bonn, Germany}

\author{M.~Thiel}
\address{Institut f\"ur Kernphysik, University of Mainz, D-55099 Mainz, Germany}

\author{A.~Thomas}
\address{Institut f\"ur Kernphysik, University of Mainz, D-55099 Mainz, Germany}

\author{M.~Unverzagt}
\address{Institut f\"ur Kernphysik, University of Mainz, D-55099 Mainz, Germany}

\author{Yu.A.~Usov}
\address{Joint Institute for Nuclear Research, 141980 Dubna, Russia}

\author{N.K.~Walford}
\address{Department of Physics, University of Basel, Ch-4056 Basel, Switzerland}

\author{D.P.~Watts}
\address{Department of Physics, University of York, Heslington, York, Y010 5DD, UK}

\author{S.~Wagner}
\address{Institut f\"ur Kernphysik, University of Mainz, D-55099 Mainz, Germany}

\author{D.~Werthm\"uller}
\address{Department of Physics, University of York, Heslington, York, Y010 5DD, UK}

\author{J.~Wettig}
\address{Institut f\"ur Kernphysik, University of Mainz, D-55099 Mainz, Germany}

\author{L.~Witthauer}
\address{Department of Physics, University of Basel, Ch-4056 Basel, Switzerland}

\author{M.~Wolfes}
\address{Institut f\"ur Kernphysik, University of Mainz, D-55099 Mainz, Germany}

\collaboration{A2 Collaboration at MAMI}
\date{\today}
\begin{abstract}
This work measured $d\sigma/d\Omega$ for neutral kaon photoproduction reactions from threshold up to a c.m.\ energy of 1855~MeV, focussing specifically on the $\gamma p\rightarrow K^0\Sigma^+$, $\gamma n\rightarrow K^0\Lambda$, and $\gamma n\rightarrow K^0 \Sigma^0$ reactions. Our results for $\gamma n\rightarrow K^0 \Sigma^0$ are the first-ever measurements for that reaction.  These data will provide insight into the properties of $N^*$ resonances and, in particular, will lead to an improved knowledge about those states that couple only weakly to the $\pi N$ channel.  Integrated cross sections were extracted by fitting the differential cross sections for each reaction as a series of Legendre polynomials and our results are compared with prior experimental results and theoretical predictions. 
\end{abstract}


\maketitle

\section{Introduction}
Most of our early knowledge of $N^*$ resonances came from experiments involving the $\pi N$ channel in the initial or final state, {\it e.g.}, pion nucleon elastic or inelastic scattering \cite{briscoe2015} or single-pion photoproduction. Lattice QCD and quark models both predict more nucleon resonances in the mass range below 2000~MeV than have been observed experimentally. This is known as the ``missing resonances'' problem in baryon spectroscopy. For that reason, there has been a concerted effort at electromagnetic facilities, including JLab, Mainz, and Bonn, to measure $N^*$ formation reactions that do not include the $\pi N$ channel at all. The data analyzed in this work bear directly on that problem.  The photoproduction of a kaon on a nucleon target can provide new information on nucleon resonances. Out of six elementary kaon photoproduction reactions ($\gamma p\rightarrow K^0\Sigma^+$, $\gamma n\rightarrow K^0\Lambda$, $\gamma n\rightarrow K^0\Sigma^0$, $\gamma p\rightarrow K^+\Lambda$, $\gamma p\rightarrow K^+\Sigma^0$, $\gamma n\rightarrow K^+\Sigma^-$), a significant amount of experimental research \cite{zegere,mcnabb,mqtran,bradford,sarantsev} has been done on the charged kaon reactions. 

By contrast, there have been very few published studies of $K^0$ photoproduction. Lawall {\it et al.} \cite{lawall} measured $\gamma p\rightarrow K^0\Sigma^+$ at ELSA, in Bonn, using the SAPHIR detector. Events were reconstructed using the $K^0\rightarrow\pi^+\pi^- $, $\Sigma^+\rightarrow \pi^0p$, and $\Sigma^+\rightarrow \pi^+n$ decays. Castelijns {\it et al.} \cite{castelijins} performed complementary measurements of $\gamma p\rightarrow K^0\Sigma^+$ at ELSA with events reconstructed using the $K^0\rightarrow\pi^0\pi^0 $ and $\Sigma^+\rightarrow \pi^0p$ decays. Aguar-Bartolom{\'e} {\it et al.} \cite{aguar} measured $\gamma p\rightarrow K^0\Sigma^+$ at Mainz using the Crystal Ball and TAPS detectors with events reconstructed using the $K^0\rightarrow\pi^0\pi^0 $ and $\Sigma^+\rightarrow \pi^0p$ decays. Recently, Compton {\it et al.} \cite{nickcompton} measured $\gamma n \rightarrow K^0 \Lambda$  at JLab using the CLAS detector. Data were collected in two datasets, g10 and g13, which used different run conditions. Events were reconstructed using the $K^0\rightarrow\pi^+\pi^- $ and $\Lambda\rightarrow\pi^-p$ decays. 

The main focus of the current work was to measure the differential cross section from threshold to c.m.~energy $W=1855$~MeV for the reactions $\gamma p \rightarrow K^0 \Sigma^+$, $\gamma n \rightarrow K^0 \Lambda$, and $\gamma n \rightarrow K^0 \Sigma^0$ on a liquid deuterium target, where $W$ was calculated from the incident beam energy assuming quasifree kinematics. The measurements were performed at MAMI-C, the Mainz Microtron located in Mainz, Germany.  We analyzed these reactions via the $K^0\rightarrow\pi^0\pi^0 $ decay. Further details are provided in Sec.~\ref{sec:three}.

The cross-section data can be used to help determine $N^{*}$ resonance properties using partial-wave analyses or to test phenomenological models of kaon photoproduction. This paper reports the world's first results on differential and total cross sections for the reaction $\gamma n\rightarrow K^0\Sigma^0$. 

This paper is divided into six sections: Sec.~\ref{sec:two} describes the experimental setup, Sec.~\ref{sec:three} describes the data analysis, Sec.~\ref{sec:four} describes the calculation of uncertainties, Sec.~\ref{sec:five} describes the results and discussion for all three reactions, and Sec.~\ref{sec:six} gives the summary and conclusions. Our measured cross sections are tabulated in the appendix.

\section{\label{sec:two}Experimental setup}
Data for the photoproduction of neutral kaon reactions on a liquid deuterium target were measured using the Crystal Ball (CB) \cite{starostin,dominik,prakhov,mcnicoll,dieterle}, particle identification detector (PID) \cite{watts2004} and TAPS \cite{prakhov,mcnicoll,dieterle} detectors. All these detectors were set up at the Mainz Microtron \cite{kaiser} bremsstrahlung-tagged photon beam facility in Germany. At the time the measurements were performed, MAMI-C could deliver electrons with energies up to a maximum energy of 1508~MeV. The mono-energetic electron beam was used to produce photons via  bremsstrahlung in a 10-$\mu$m copper radiator. The bremsstrahlung photons are tagged by the Glasgow photon tagger \cite{mcgeorge}. The tagged photons are then passed through a lead collimator to produce a photon beam. The hole in the lead collimator was 4~mm in diameter for this experiment. This collimation gave a photon beam spot on target with a diameter of about 1.3~cm.  The photon beam was incident on a 125~$\mu$m Kapton target cylinder of length 4.72~cm and diameter 4~cm.  Further details on the target system can be found in Ref.~\cite{bantawa}.

The Crystal Ball (CB) is a multiphoton spherical spectrometer \cite{starostin}. The CB geometry is based on an icosahedron, a polyhedron having 20 triangle-shaped sides. Each of the 20 major triangles is divided into four minor triangles. Each minor triangle consists of nine crystals, so for a complete sphere, there would be 720 crystals. However, for the entrance and exit tunnels, 48 crystals were not installed, resulting in 672 crystals for the Crystal Ball. The chemical composition of each crystal is thallium-doped sodium iodide, NaI(Tl), which is a hygroscopic material so it is important to protect the crystals from moisture \cite{dominik,werthmuller}. The Crystal Ball covers the polar angle range from $20^\circ$ to $160^\circ$ and the azimuthal angle range from $0^\circ$ to $360^\circ$. 

The forward moving particles are detected by TAPS \cite{dominik,Witthauer}, which was configured as a photon calorimeter consisting of 384 $\rm BaF_2$ crystals located downstream of the Crystal Ball. These $\rm BaF_2$ crystals were arranged in a honeycomb pattern to form a hexagonal wall covering the polar angle range from $4^\circ$ to $20^\circ$. 

\begin{figure}[hb]
\centering
\includegraphics[scale=.60]{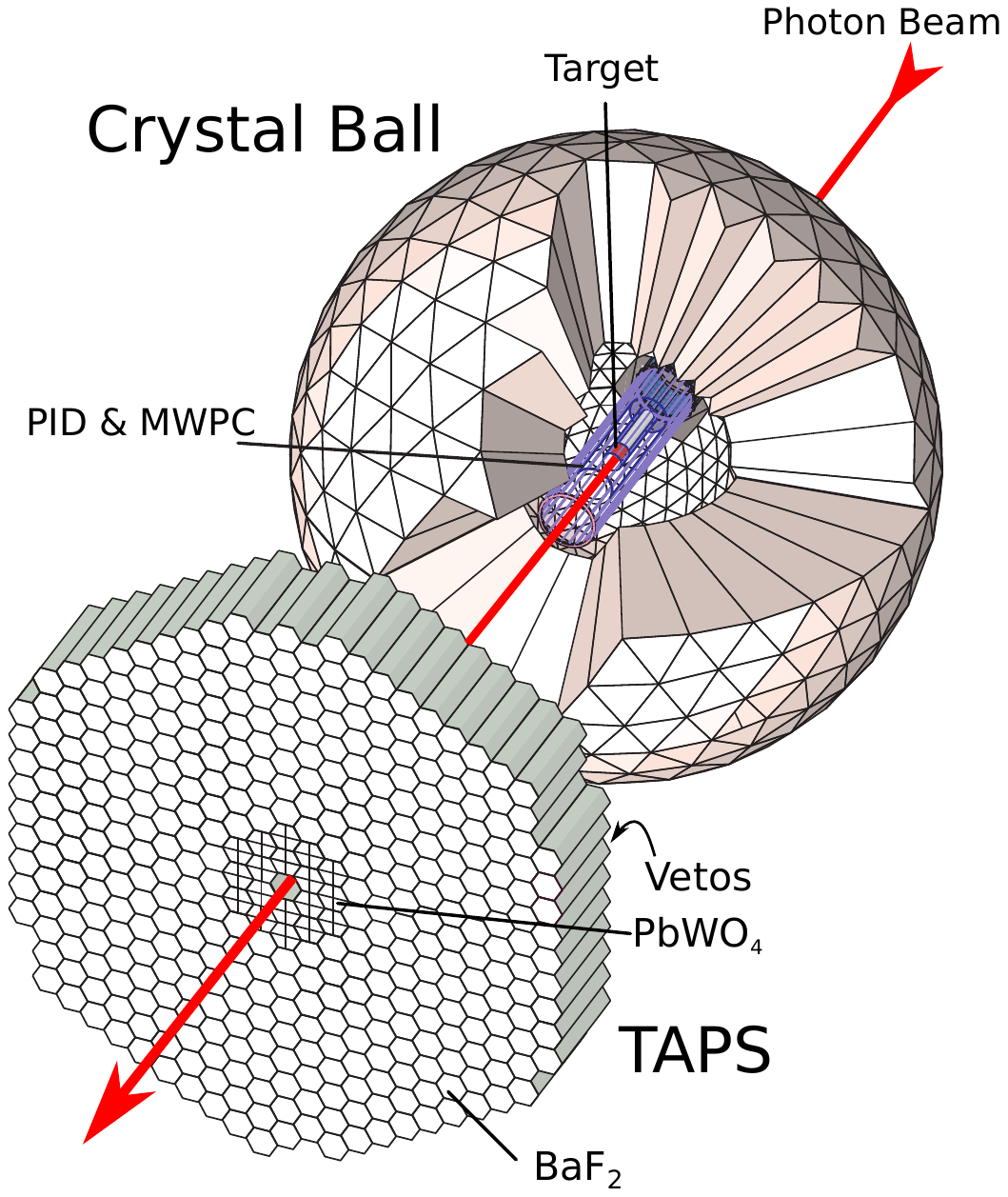}

\caption{Schematic diagram of the CB and TAPS detectors. The PID is placed inside the CB for charged particle detection. In this experiment, the PbWO$_4$ crystals were not installed in TAPS.}
\label{CBTAPS}
\end{figure}

The PID (Particle Identification Detector) \cite{watts2004} is a cylindrical detector with a 5-cm inner radius oriented concentric with the target inside the Crystal Ball. It was designed to work along with the CB to provide information on charged particles. The PID distinguishes between different types of charged particles and neutral particles based on the energy deposited in the PID elements versus total energy measured in a CB cluster. For further details about these detectors, such as their energy and angle resolutions or their calibrations, see \cite{zehr,prakhov,mcnicoll,werthmuller, kaser, oberle, schumann, dieterle}. The CB and TAPS detectors are very efficient at detecting the final-state photons. A cylindrical MWPC (MultiWire Proportional Chamber) may be used to improve the angular resolution (tracking) of charged particles. During this experiment, the MWPC was not used. Figure~\ref{CBTAPS} shows a schematic diagram of the CB and TAPS detector setup.

\section{\label{sec:three}Data Analysis}
After all the detectors had been calibrated, the event selection and analysis was carried out.  Detailed Monte Carlo (MC) studies were performed using $3 \times 10^6$ events generated according to phase space for each of the three $K^0$ photoproduction reactions, as well as for $\gamma p \to \eta p$ and $\gamma n \to \eta n$, which are the leading backround reactions due to $\eta \to 3\pi^0 \to 6\gamma$ decays. 

 In each reaction the $K^0$ was identified through its decay $K^0\rightarrow \pi^0\pi^0\rightarrow 4\gamma$.  The $\Sigma^+$ was identified through its decay $\Sigma^+\rightarrow \pi^0p$, $\Lambda$ through its decay $\Lambda\rightarrow \pi^0n$, and $\Sigma^0$ through its decay $\Sigma^0\rightarrow \gamma\Lambda\rightarrow\gamma\pi^0n$. Therefore, the detection of three $\pi^0$s in the final state was required in all cases, giving rise to six final-state photons via $\pi^0\rightarrow\gamma\gamma$. Data for $\gamma p\rightarrow K^0\Sigma^+$, $\gamma n\rightarrow K^0\Lambda$, and $\gamma n\rightarrow K^0\Sigma^0$ reactions were sorted into various cases $(nc)$, where $n$ represents the detected number of final-state neutral particles and $c$ represents the detected number of final-state charged particles. Table~\ref{differentcases} tabulates the reactions and the corresponding cases for the present work.
\begin{table}[H]
\caption{Cases based on nucleon detection for all three $\gamma N\rightarrow K^0Y$.}
\begin{ruledtabular}
\begin{tabular}{ccc}
				Case	& Reaction & Comment \\   				\hline
$61$&$\gamma p\rightarrow K^0\Sigma^+$&final $p$ detected\\ \hline
$60$&$\gamma p\rightarrow K^0\Sigma^+$&final $p$ not detected\\
&$\gamma n\rightarrow K^0\Lambda$&final $n$ not detected\\ \hline
$70$&$\gamma n\rightarrow K^0\Lambda$&final $n$ detected\\
&$\gamma n\rightarrow K^0\Sigma^0$&final $n$ not detected\\ \hline
$80$&$\gamma n\rightarrow K^0\Sigma^0$&final $n$ detected
\label{differentcases}
\end{tabular}
\end{ruledtabular}
\end{table}		
If only six neutral clusters are detected, the event is case (60). To be a viable event for $\gamma p\rightarrow K^0\Sigma^+$ or $\gamma n\rightarrow K^0\Lambda$, further analysis was needed to establish these six neutral clusters as photons produced from $\pi^0$ decays. The data analysis for case (60) starts by first selecting events that have six and only six neutral clusters. If the final proton in $\Sigma^+\rightarrow \pi^0p$ is detected then there will be six neutral clusters and one charged cluster in the final state, which defines case (61). If the neutron in $\Lambda\rightarrow \pi^0n$ is detected then there will be seven neutral clusters and no charged cluster, which defines case (70). 

For $\gamma n\rightarrow K^0\Sigma^0$ events, the detection of seven photon candidates is required, six coming from $\pi^0$ decays and one coming from $\Sigma^0\rightarrow \gamma\Lambda$. If the final-state neutron is not detected, then the event corresponds to case (70); however, if the final-state neutron is detected, then the event corresponds to case (80).

\begin{figure}[tp]
\centering
\includegraphics[scale=.47]{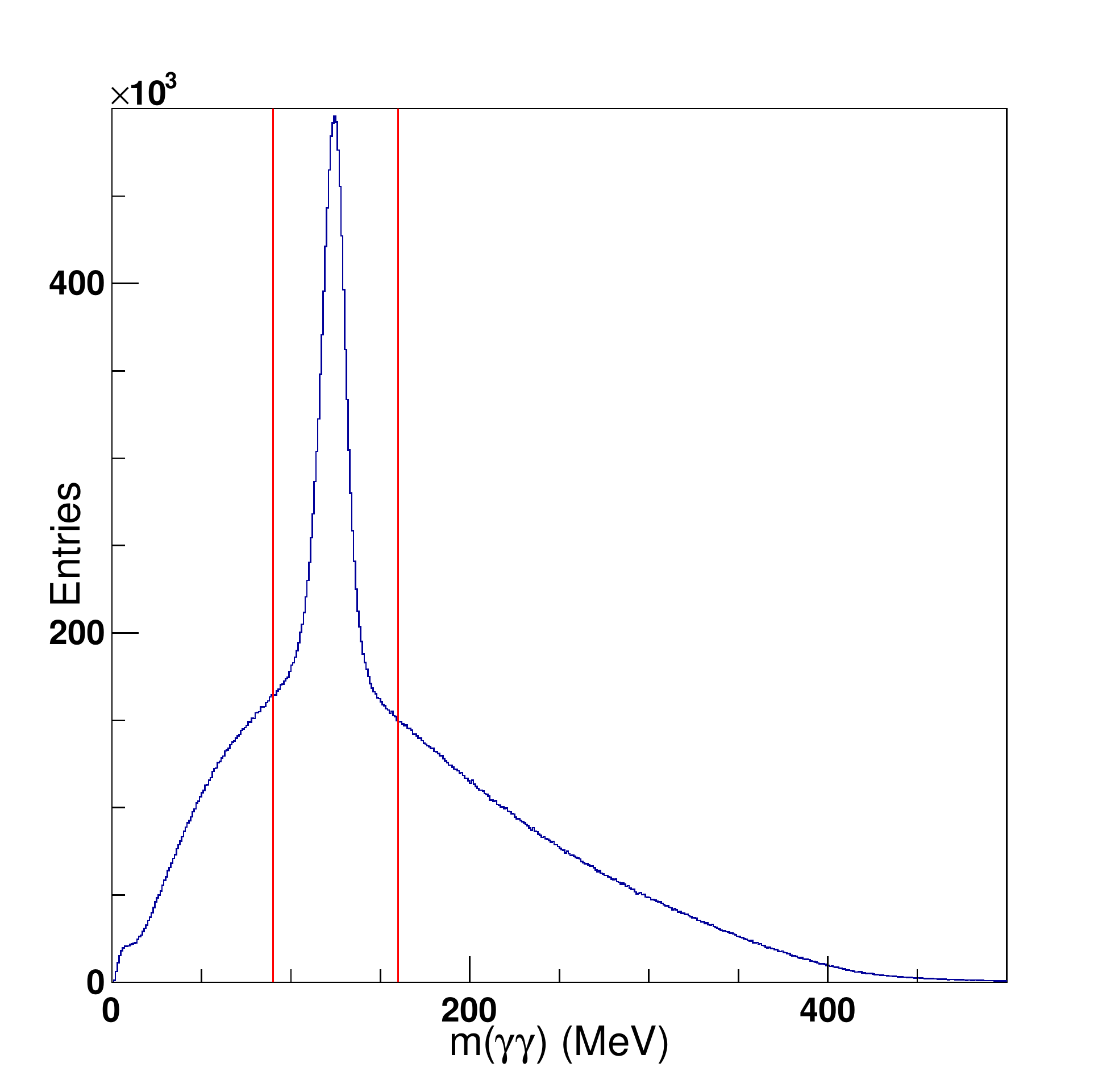}
\caption{Invariant mass of all distinct $\gamma\gamma$ combinations for Monte Carlo simulated $\gamma n\rightarrow K^0\Lambda$ events for case (60). The peak corresponds to the $\pi^0$ meson. Combinations between the cuts denoted by the vertical red lines correspond to $\pi^0$ candidates.}
\label{im2g}
\end{figure}

\begin{figure}[t]
\centering
\includegraphics[scale=.43]{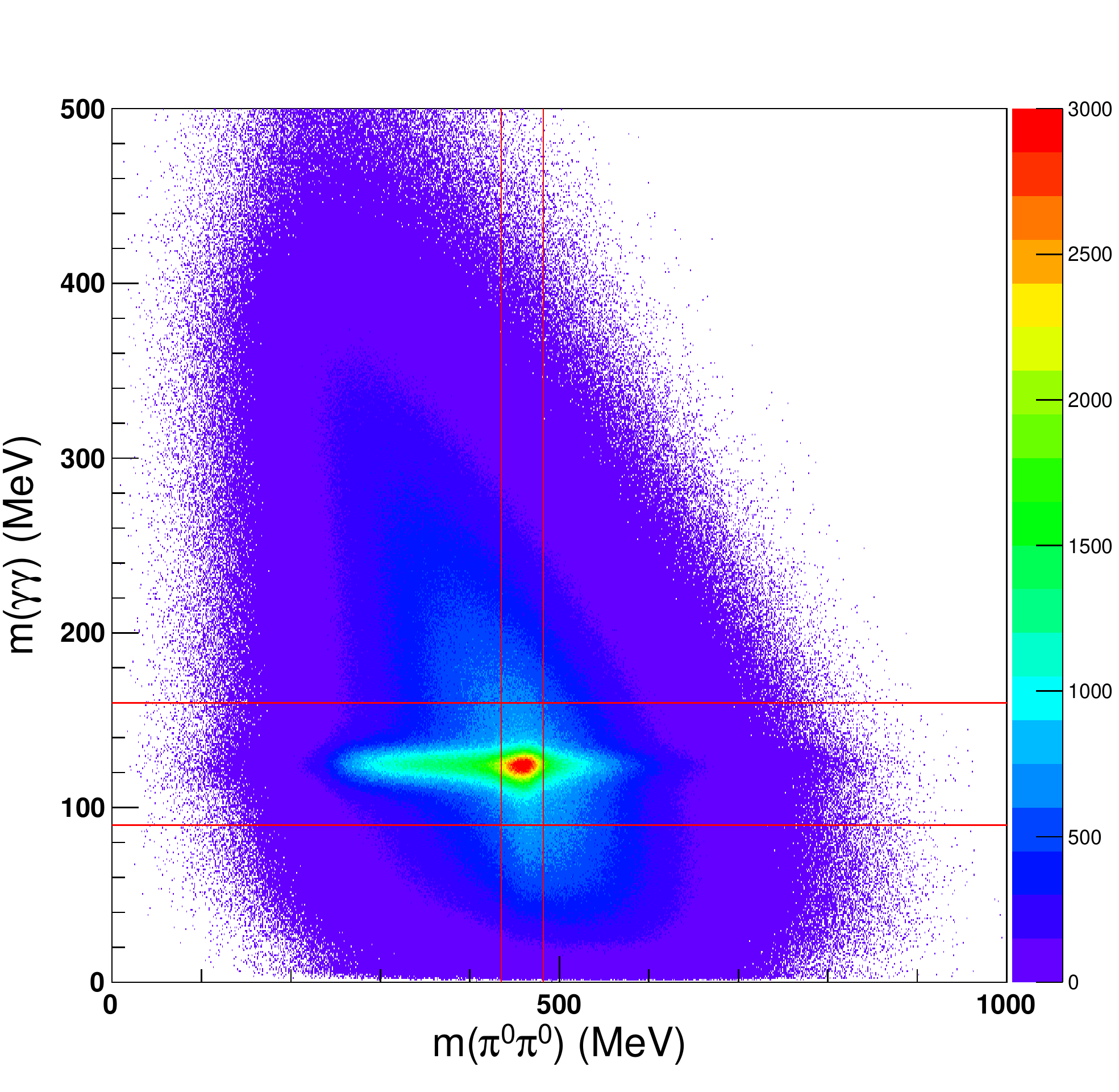}
\caption{Invariant mass $m$($\pi^0\pi^0$) vs. invariant mass $m(\gamma\gamma)$ for Monte Carlo simulated $\gamma n\rightarrow K^0\Lambda$ events for case (60). The photon candidates used to calculate $m(\gamma\gamma)$ were distinct from those used to calculate $m(\pi^0\pi^0)$.}
\label{im2gvsim4g}
\end{figure}

Once events had been separated according to the number of neutral and charged clusters, the next step was to identify the final three $\pi^0$s from the neutral clusters. To identify the three $\pi^0$s, all distinct possible combinations of two-photon candidates were constructed. There are 15, 21, and 28 possible ways to construct distinct two-$\gamma$ combinations from six, seven, and eight neutral clusters, respectively. A histogram of the invariant-mass of all distinct two-$\gamma$ combinations for case (60) is shown in Fig.~\ref{im2g}. Only those  distinct two-$\gamma$ combinations whose invariant-mass $m(\gamma\gamma$) was between 90 and 160~MeV are the actual $\pi^0$ candidates.  This invariant-mass cut is represented by solid red vertical lines in Fig.~\ref{im2g}. 
A typical event had several combinations that satisfied this criterion. Only those events that had a minimum of three distinct $\pi^0$ candidates were kept. Major sources of background for the reactions of interest are $\gamma p\rightarrow \eta p$ and $\gamma n\rightarrow \eta n$, where $\eta\rightarrow3\pi^0$. In order to eliminate this background, only those three $\pi^0$ candidates whose combined invariant mass is greater than 600~MeV were selected for further analysis \cite{aguar,nanova}. This cut significantly reduces the $\eta$ background contribution while only slightly reducing events from the reactions of interest. If the three $\pi^0$ candidates for a given combination are labeled as $\pi^0_{1}$, $\pi^0_{2}$, $\pi^0_{3}$, then there are three ways to construct the two $\pi^0$s that could correspond to a $K^0$ decay; that is, ($\pi^0_{1}\pi^0_{2}$), ($\pi^0_{2}\pi^0_{3}$), or ($\pi^0_{1}\pi^0_{3}$). A histogram of the mass of one $\pi^0$ candidate $m(\gamma\gamma)$ versus the invariant mass  m($\pi^0\pi^0$) of the other two $\pi^0$ candidates is shown in Fig.~\ref{im2gvsim4g}. This two-dimensional plot provided information on where best to impose a cut on $m(\pi^0\pi^0)$ to reduce the background further. Only combinations in which $m(\pi^0\pi^0)$ was between 435 and 482~MeV were selected for further analysis. This cut was applied before the energy correction discussed below.  After this correction, the $K^0$ peaks in the $\pi^0\pi^0$ invariant-mass distribution were very close to 498~MeV.

\begin{figure}[h]
\centering
\includegraphics[scale=.43]{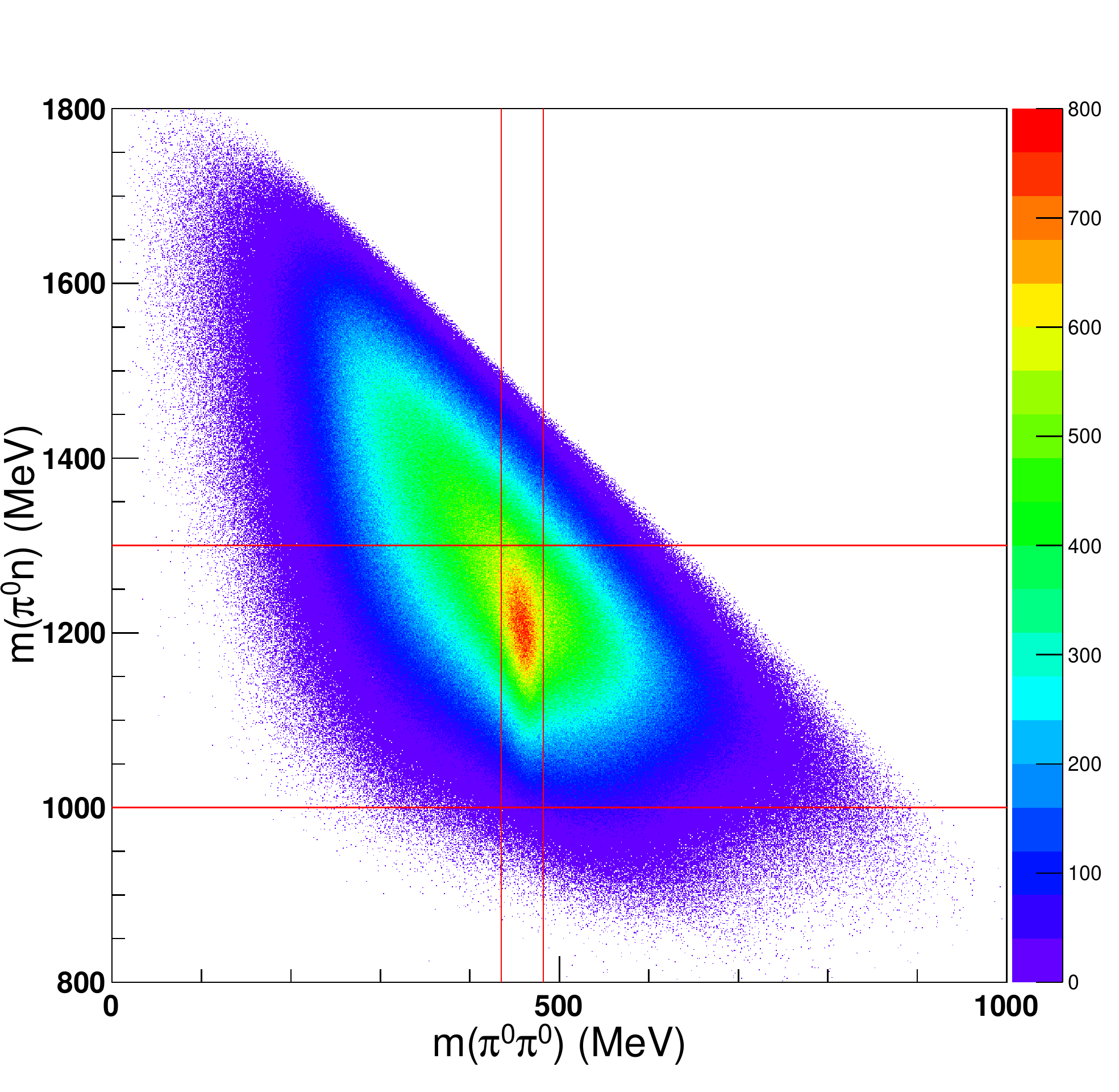}
\caption{Invariant mass $m$($\pi^0n$) vs. invariant mass $m$($\pi^0\pi^0$) for Monte Carlo simulated $\gamma n\rightarrow K^0\Lambda$ events for case (60).  (See text for details.)}
\label{im4gvsmm4g}
\end{figure}

Figure~\ref{im4gvsmm4g} 
shows a histogram of the invariant mass $m(\pi^0n)$ plotted versus the invariant mass $m(\pi^0\pi^0)$. The quantity $m(\pi^0n)$ was actually calculated as the missing mass of the same $\pi^0\pi^0$ combination, since the two quantities should be equal.  This plot provided information on where best to impose a cut on the invariant mass $m(\pi^0n)$. Only combinations in which $m(\pi^0n)$ was between 1000 and 1300~MeV were selected for further analysis.  After the energy correction mentioned above and described below, the peaks in the $m(\pi^0n)$ distributions were very close to the $\Lambda$ mass (1116~MeV) for the MC simulated $\gamma n \to K^0 \Lambda$ events. Monte Carlo studies on the polar angle of the undetected nucleon showed that most of the undetected nucleons go forward at our kinematics. A cut was therefore imposed that the cosine of the polar angle of the final-state nucleon, whether measured or calculated, must be greater than or equal to 0.7. All these cuts were used to reduce the number of incorrect three $\pi^0$ combinations. Even after all these cuts, there were still a number of  events with more than one candidate for the correct three-$\pi^0$ combination. Monte Carlo studies were made of the opening angle between two photons for $\pi^0\rightarrow\gamma\gamma$ decays. While the distribution is broad, it is more likely for our kinematics that the opening angle is less than $90^\circ$ than greater than $90^\circ$. The average opening angle for each remaining three $\pi^0$ combination was therefore calculated and the combination with the minimum average opening angle was selected as the best choice for the correct 3$\pi^0$ combination.  Although several methods for reconstructing the 3$\pi^0$ combination  were investigated using Monte Carlo simulations, this method produced the largest $K^0$ yields.

For case (61), events with six neutral clusters and one charged cluster were selected. The PID was used to select the proton candidate. Similar analysis steps were used to select the best choice for the correct three-$\pi^0$ combination as for case (60). 

For case (70), there were seven neutral clusters. Similar analysis steps were followed as for case (60) to identify the best choice for the correct three-$\pi^0$ combination. Here for each three-$\pi^0$ combination there was one unpaired particle.

For case (80), there were eight neutral clusters. Again, similar analysis steps were followed as for case (60) to identify the best choice for the correct three $\pi^0$ combination. Here for each three-$\pi^0$ combination there were two unpaired particles; i.e, the seventh and eighth particles (a photon and a neutron). The missing mass of the seven photons should equal the mass of the neutron. Therefore a cut was imposed that the missing mass of the three $\pi^0$s and the seventh particle (a photon) be greater than 800~MeV and a cut that cosine of the polar angle of the eighth particle (a neutron) should be greater than or equal to 0.7. These cuts were used to reduce the number of incorrect three-$\pi^0$ combinations for case (80), and helped to distinguish which of the other neutral particles was a neutron. 

The energy reconstruction of the $K^0$ mesons was improved by applying a correction \cite{werthmuller},
\begin{equation} 
\label{scalling}
E^{'}=E\cdot\frac{m_{\pi^0}}{m_{\gamma\gamma}},
\end{equation} 
which made use of information obtained from the good angular resolution of the CB, after the best choice for the correct three-$\pi^0$ combination had been determined. Here $E$ is the relativistic energy of each $\pi^0$, $m_{\gamma\gamma}$ is the invariant mass of the decay photons, and $m_{\pi^0}$ = 135~MeV is the known $\pi^0$ mass. Before scaling, the invariant mass for $\pi^0\rightarrow\gamma_{1}\gamma_{2}$ is given by 
\begin{equation} 
{(m_{\gamma\gamma})^2}=2E_{1}E_{2}{(1-\cos\theta_{\gamma\gamma})},
\end{equation}
where $\theta_{\gamma\gamma}$ is the measured opening angle for $\pi^0\rightarrow\gamma_{1}\gamma_{2}$. Here $E_{1}$ and $E_{2}$ are the measured energies of the two photon clusters.
After scaling ($E_1 \rightarrow E^{'}_{1}$ and $E_2 \rightarrow E^{'}_{2}$), the scaled invariant mass $m(\gamma\gamma)$ was exactly the $\pi^0$ mass, 135~MeV. The scaled 4-momenta of the $\pi^0$s were used to calculate $m(\pi^0\pi^0)$ and $m(\pi^0N)$, where $N$ represents the nucleon. All three $\pi^0\pi^0$ combinations were considered for further analysis. In MC simulations for each $\gamma N\rightarrow K^0 Y$ event, there are two incorrect $\pi^0\pi^0$ combinations for every correct combination corresponding to $K^0\rightarrow\pi^0\pi^0$. In real data, there can be additional contributions to background in the $m(\pi^0\pi^0)$ distributions.

The $\pi^{0}\pi^{0}$ invariant-mass distributions were fitted using a binned likelihood method with the parametrization
\begin{eqnarray}
y(x) &=& \Big[\frac{x^2-(270)^2}{x^2}\Big]^{\alpha}\Big[\beta \exp\Big(-{\frac{1}{2}\Big(\frac{x-\mu}{\sigma_{B}}\Big)^2}\Big)\nonumber \\ 
&+& \delta  \exp\Big(-{\frac{1}{2}\Big(\frac{x-498}{\sigma_{K}}\Big)^2}\Big)\Big],
\end{eqnarray}
where $\alpha$, $\beta$, $\delta$, $\mu$, $\sigma_{B}$, and $\sigma_{K}$ were fitting parameters. The first factor ensured that the distribution goes to zero when $x=2m_{\pi^0}=270$~MeV. The exponent $\alpha$ is a small number ($0<\alpha<1$) determined by fitting the $m(\pi^0\pi^0)$ distribution for given energy bins. The parameter $\beta$ measures the yield of the background contribution. The background was represented by a scaled Gaussian distribution with centroid $\mu$ and standard deviation $\sigma_{B}$. The parameter $\delta$ measures the yield of the kaon signal.  The kaon signal distribution was represented by a scaled Gaussian with centroid 498~MeV (the $K^0$ mass) and standard deviation $\sigma_{K}$. The observed $m(\pi^0\pi^0)$ distributions for each energy bin, summed over all angle bins, were fitted to determine $\alpha$ and $\sigma_{K}$ parameter values for each energy bin. Next the observed $m(\pi^0\pi^0)$ distributions for each angle bin, for a particular energy bin, were fitted with the values of $\alpha$ and $\sigma_{K}$ held fixed at their fitted values for that particular energy bin. Monte Carlo simulations were used to verify that this approximation was reasonable.  The fitting parameters $\beta$, $\delta$, $\mu$, and $\sigma_{B}$ were allowed to vary freely in each angle and energy bin. The fitted value of $\mu$ for a particular angle and energy bin, with $\alpha$ and $\sigma_{K}$ held fixed as described above, was called the nominal background centroid. The values of the nominal background centroid for each energy and angle bin were recorded for further analysis. The background contribution was obtained after the fit by setting $\delta$ equal to zero. Numerical integration was used to calculate the total number of kaons (the kaon yield, $N_{K^0}$) by subtracting the areas under the total and background curves. 

\begin{figure*}
	\includegraphics[scale=0.3]{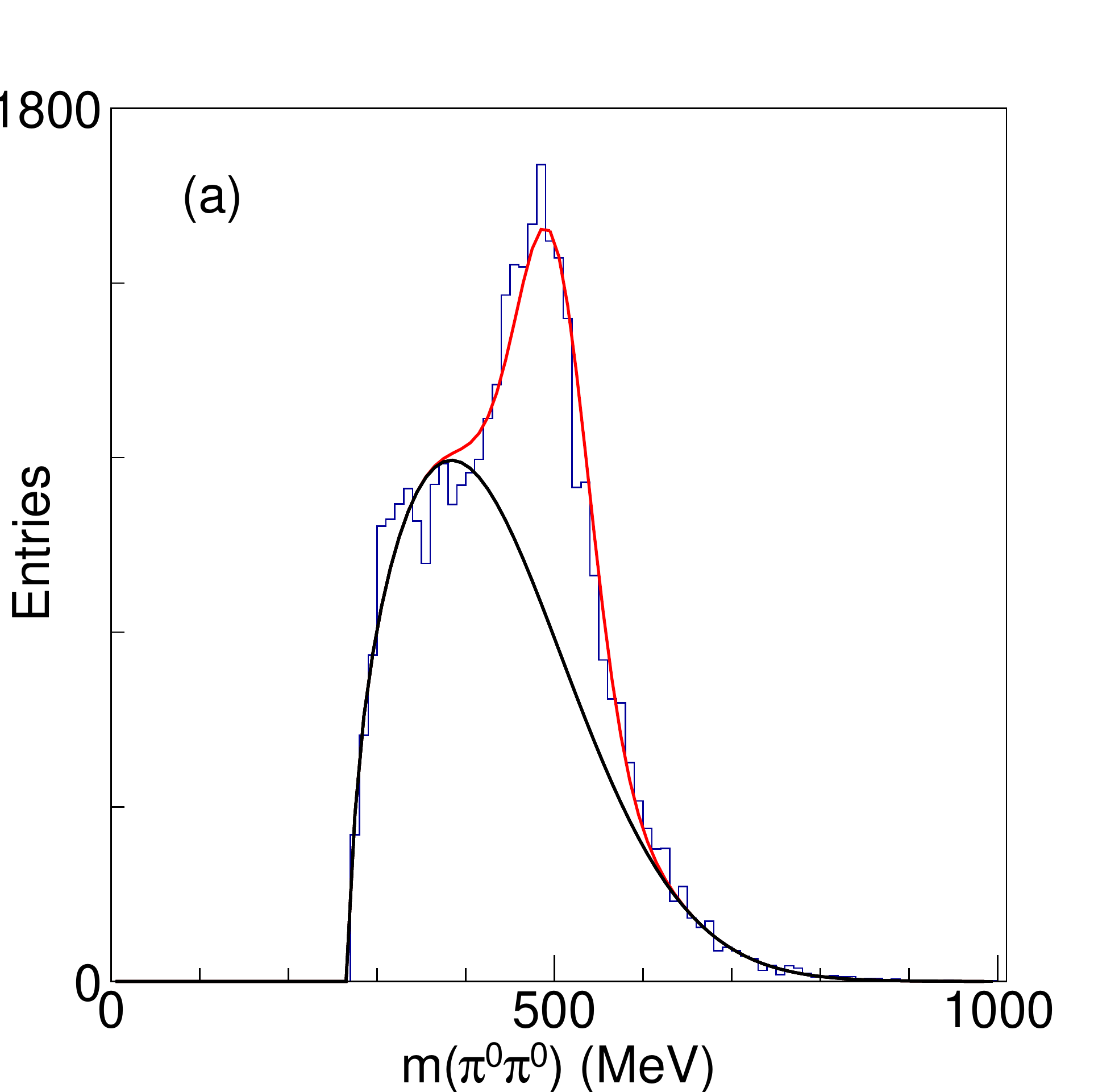}
	\includegraphics[scale=0.3]{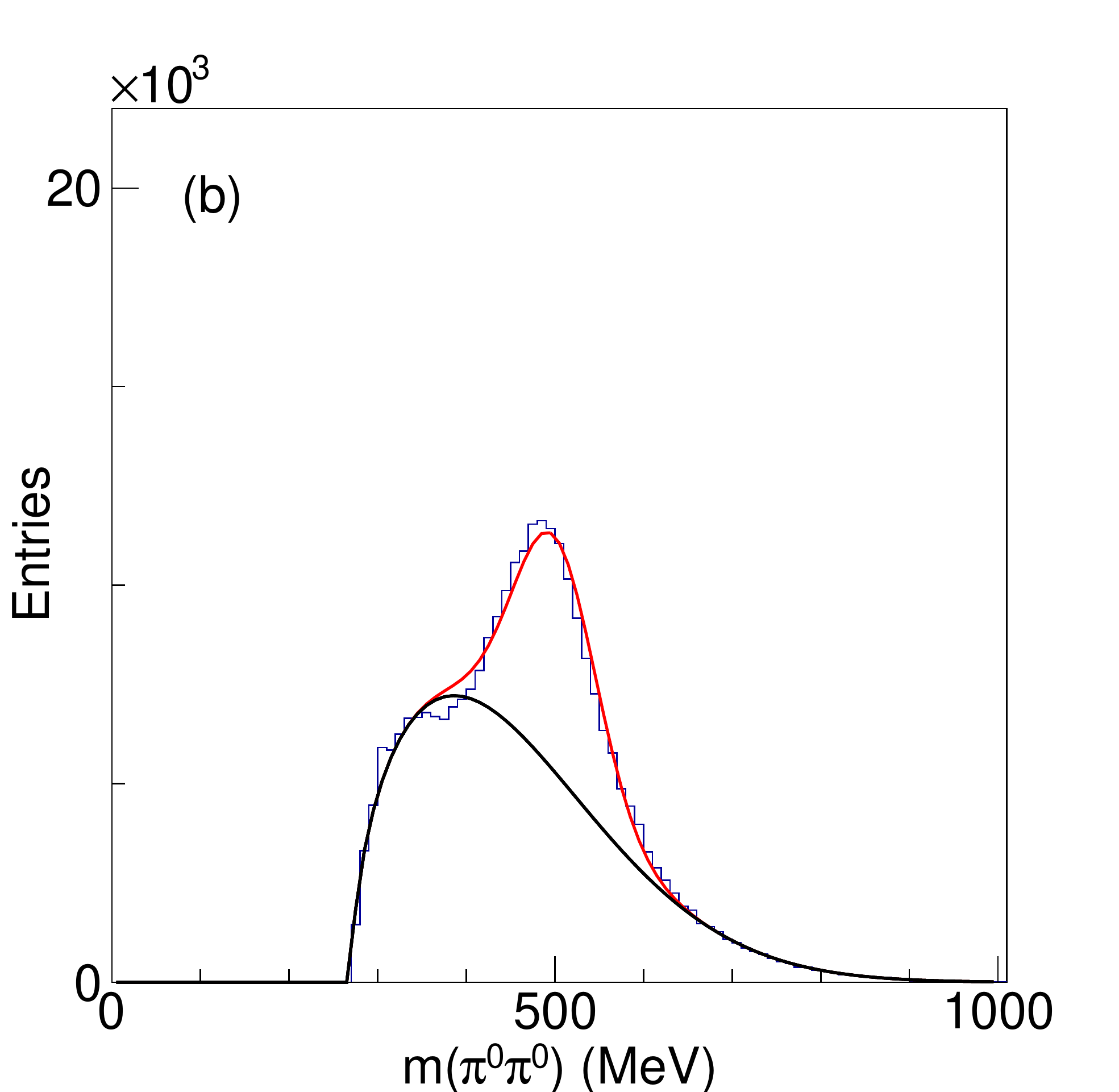}
	\includegraphics[scale=0.3]{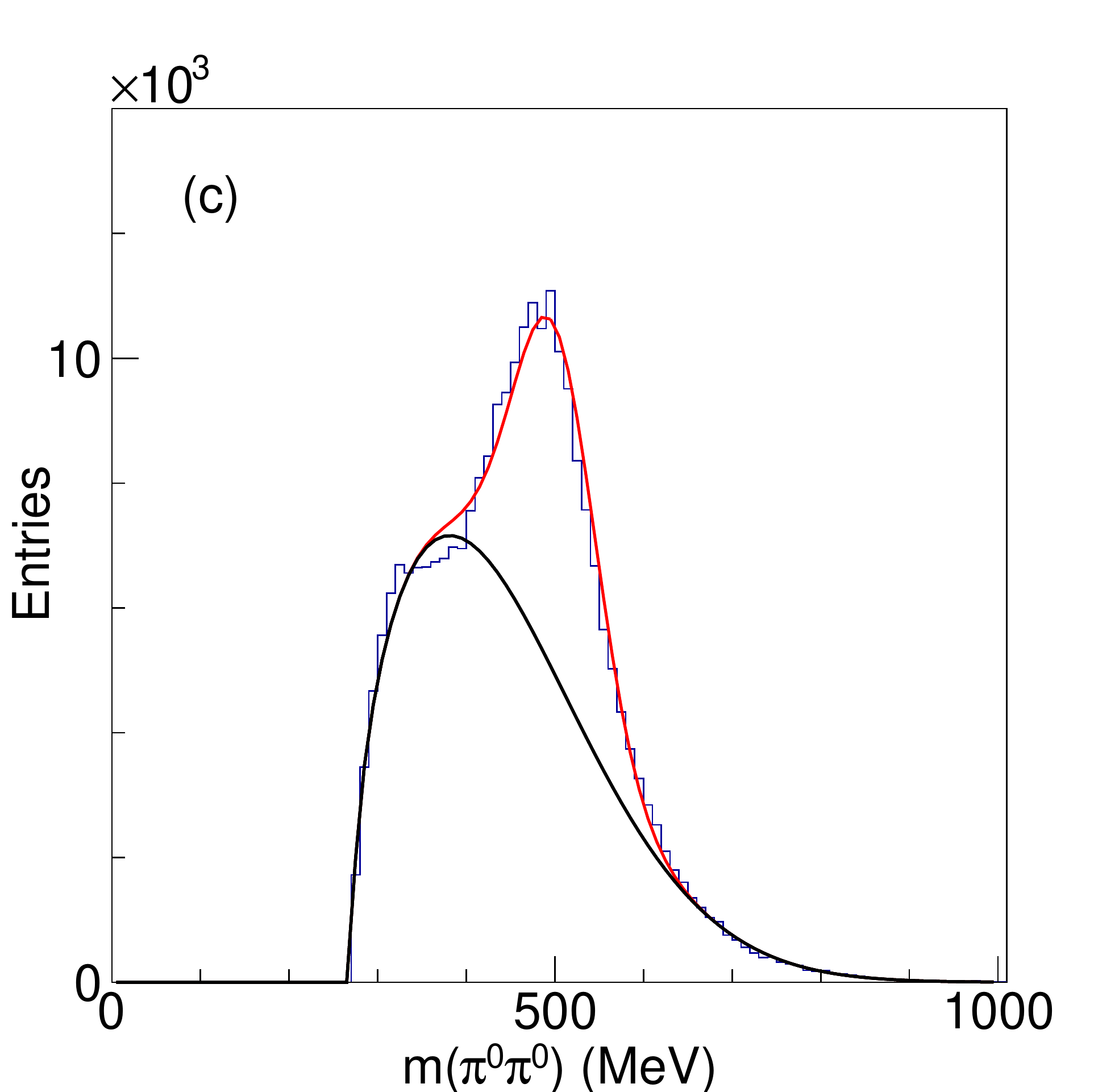}
	\includegraphics[scale=0.3]{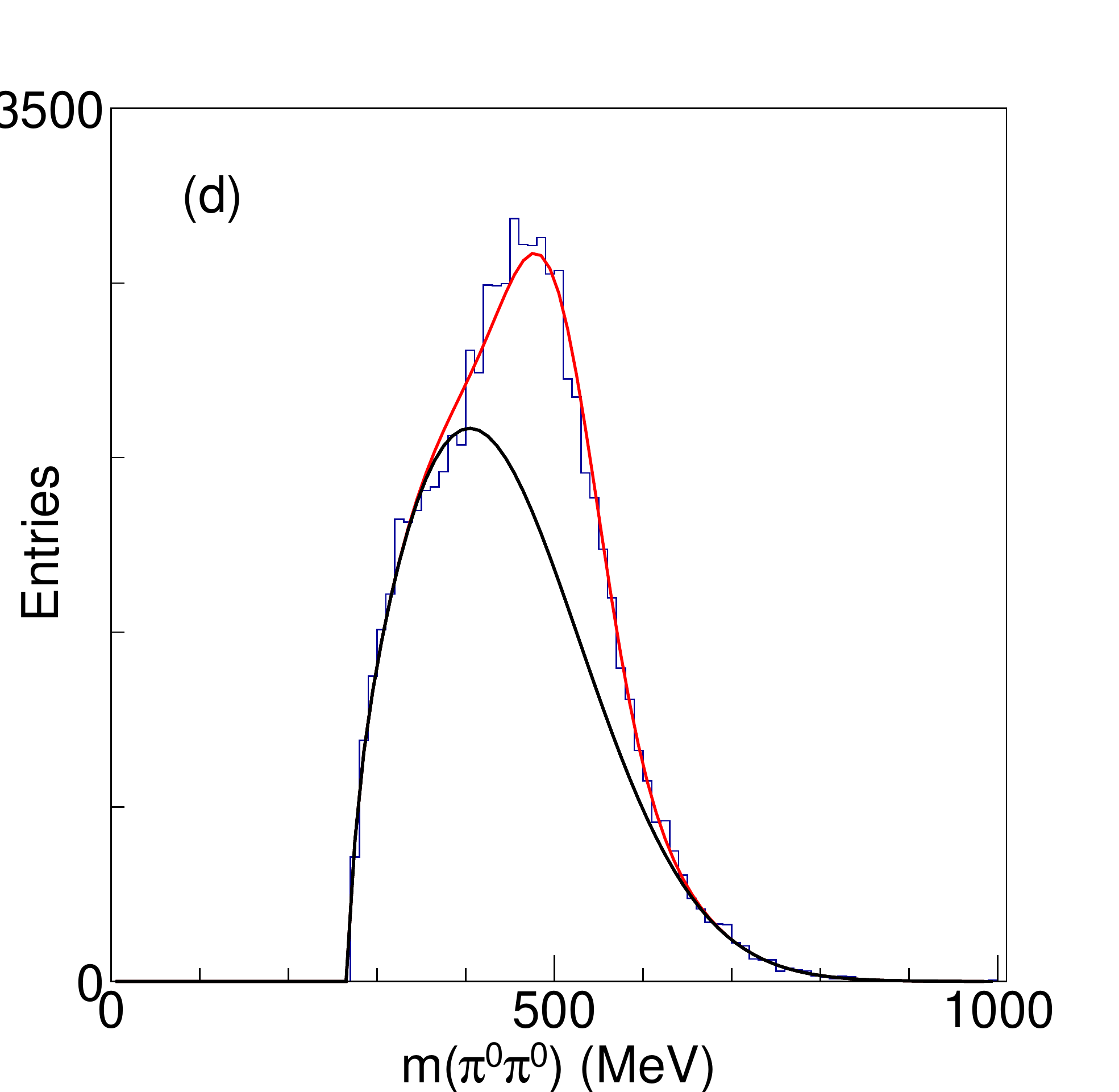}
			\caption{Observed $\pi^0\pi^0$ invariant-mass distributions: (a) $\gamma p\rightarrow K^0 \Sigma^+$ for  case 61, (b) combination of $\gamma p\rightarrow K^0 \Sigma^+$ and $\gamma n\rightarrow K^0 \Lambda$ for case 60, (c) combination of $\gamma n\rightarrow K^0 \Lambda$ and $\gamma n\rightarrow K^0 \Sigma^0$ for case 70, (d) $\gamma n\rightarrow K^0 \Sigma^0$ for  case 80. Data in the histograms were summed over all energy and angle bins. The fitted total invariant-mass distributions are represented by solid red curves and the background contributions are represented by solid black curves.}	
	\label{realinvariantmass}
\end{figure*}

The kaon yield was sensitive to the background contribution.  A second fit of the observed $m(\pi^0\pi^0)$ distributions was performed with a different value of $\mu$ called the modified centroid. The modified centroid was chosen to be the average of the nominal centroid of the background and the signal centroid (498~MeV). This modified centroid was the maximum value of the background centroid that produced a good fit of the data. The use of these two background centroids is discussed further in Sec.~\ref{sec:four}.  Figure~\ref{realinvariantmass} 
shows the observed $\pi^{0}\pi^{0}$ invariant-mass distributions for $\gamma p\rightarrow K^0 \Sigma^+$, $\gamma n\rightarrow K^0 \Lambda$, and $\gamma n\rightarrow K^0 \Sigma^0$ summed over all energy and angle bins. The fitted total invariant-mass distributions are represented by solid red curves and the background contributions are represented by solid black curves. 

For calculating the differential cross sections, eight angle bins were used to cover the range from $\cos\theta_{\rm cm}$= $-$1.0 to $+$1.0. The c.m.\ energy range $W=1615$ to $1765$~MeV was divided into five bins of width 30~MeV, and the c.m.~energy range $W=1765$ to $1865$~MeV was divided into five bins of width 20~MeV. After subtracting the background, the differential cross section for a specified energy-angle bin was calculated using   
\begin{equation} 
\label{diffcrossform}
\frac{d\sigma}{d\Omega} = \frac{\it N_{K^{0}}} {\it N_{\gamma}\epsilon \it N_{\rm t} {\it B} \rm 2\pi \Delta \cos\theta_{\rm cm}} ,
\end{equation}
where $N_{K^{0}} = N_{K^{0}}(E_{\gamma},\theta_{\rm cm})$ is the kaon yield for a given energy-angle bin, $N_{\gamma} = N_{\gamma} (E_{\gamma})$ is the photon flux for a given energy bin, $\epsilon = \epsilon(E_{\gamma},\theta_{\rm cm})$ is the acceptance for a specified energy-angle bin calculated from Monte Carlo simulations, ${N_{\rm t}}$ is the number of target nucleons per ${\rm cm^{2}}$, $B$ is a product of branching ratios for the particular reaction, and $\rm{\Delta \cos\theta_{\rm cm}}$ is the bin width for  $\rm{\cos\theta_{\rm cm}}$.

The differential cross section for $\gamma n \rightarrow K^0 \Sigma^{0}$ for case (80) was calculated using 
\begin{equation}  
\label{kscase80diffcross}
\Big(\frac{d\sigma}{d\Omega}\Big)^{80}_{\gamma n \rightarrow K^0 \Sigma^0}= \frac{N^{80}_{K^{0}}}{{N_{\gamma}\epsilon^{80}_{\Sigma^{0}}N_{\rm t}{B_{\Sigma^{0}}}}2\pi \Delta \cos\theta_{\rm cm}},\\
\end{equation}
where $N^{80}_{K^0}$ is the measured $K^0$ yield for case (80) and $B_{\Sigma^{0}} = 0.05301 \pm 0.00074$. 

For case (70), the measured $K^0$ yield has contributions from both $\gamma n \rightarrow K^0 \Sigma^0$ and $\gamma n \rightarrow K^0 \Lambda$: $N^{70}_{K^0} =N^{70}_{\Lambda}+N^{70}_{\Sigma^0}$. Since $(d\sigma/d\Omega)^{70}_{\gamma n \rightarrow K^0 \Sigma^0}$ = $(d\sigma/d\Omega)^{80}_{\gamma n \rightarrow K^0 \Sigma^0}$, 
\begin{equation} 
\begin{split}
N^{70}_{K^{0}} &= \Big(\frac{d\sigma}{d\Omega}\Big)^{70}_{\gamma n \rightarrow K^0 \Lambda}\times{N_{\gamma}\epsilon^{70}_{\Lambda}N_{\rm t}{B_{\Lambda}}}2\pi \Delta \cos\theta_{\rm cm}\\
               & + \Big(\frac{d\sigma}{d\Omega}\Big)^{80}_{\gamma n \rightarrow K^0 \Sigma^{0}}\times{N_{\gamma}\epsilon^{70}_{\Sigma^{0}}N_{\rm t}{B_{\Sigma^{0}}}}2\pi \Delta \cos\theta_{\rm cm},
\end{split}
\end{equation}
where $B_{\Lambda} = B_{\Sigma^{0}} = 0.05301 \pm 0.00074$.  Values of $B_{\Lambda}$, $B_{\Sigma^{0}}$, and $B_{\Sigma^{+}}$ were calculated using branching ratios taken from the {\it Review of Particle Physics} \cite{pdg}.  For details, see Ref.~\cite{akondi}.  The $\gamma n \rightarrow K^0 \Lambda$ differential cross section for case (70) is then  
\begin{equation}  
\begin{split}
\Big(\frac{d\sigma}{d\Omega}\Big)^{70}_{\gamma n \rightarrow K^0 \Lambda}=
\frac{N^{70}_{K^{0}}}{{N_{\gamma}\epsilon^{70}_{\Lambda}N_{\rm t}{B_{\Lambda}}}2\pi \Delta \cos\theta_{\rm cm}}\\
  -~\frac{\epsilon^{70}_{\Sigma^{0}}}{\epsilon^{70}_{\Lambda}}\cdot\frac{B_{\Sigma^{0}}}{B_{\Lambda}}\cdot\Big(\frac{d\sigma}{d\Omega}\Big)^{80}_{\gamma n \rightarrow K^0 \Sigma^{0}}.
\end{split}
\end{equation}
The measured $\gamma n \rightarrow K^0 \Sigma^{0}$ cross sections for case (80) and the measured $K^0$ yields for case (70) were used to calculate the $\gamma n \rightarrow K^0 \Lambda$ cross sections for case (70). 

Similarly, the differential cross section for $\gamma p \rightarrow K^0 \Sigma^{+}$ for case (61) was calculated using 
\begin{equation}  
\label{kscase61diffcross}
\begin{split}
\Big(\frac{d\sigma}{d\Omega}\Big)^{61}_{\gamma p \rightarrow K^0 \Sigma^+}= \frac{N^{61}_{K^{0}}}{{N_{\gamma}\epsilon^{61}_{\Sigma^{+}}N_{\rm t}{B_{\Sigma^{+}}}}2\pi \Delta \cos\theta_{\rm cm}},\\
\end{split}
\end{equation}
where $N^{61}_{K^0}$ is the measured $K^0$ yield for case (61) and $B_{\Sigma^{+}} = 0.07637 \pm 0.00046$.

For case (60), the measured $K^0$ yield has contributions from both $\gamma n \rightarrow K^0 \Lambda$ and $\gamma p \rightarrow K^0 \Sigma^+$: $N^{60}_{K^0} = N^{60}_{\Sigma^+} + N^{60}_{\Lambda}$. Since $(d\sigma/d\Omega)^{60}_{\gamma p \rightarrow K^0 \Sigma^+}$ = $(d\sigma/d\Omega)^{61}_{\gamma p \rightarrow K^0 \Sigma^+}$, 
\begin{equation} 
\begin{split}
N^{60}_{K^{0}} &= \Big(\frac{d\sigma}{d\Omega}\Big)^{60}_{\gamma n \rightarrow K^0 \Lambda}\times{N_{\gamma}\epsilon^{60}_{\Lambda}N_{\rm t}{B_{\Lambda}}}2\pi \Delta \cos\theta_{\rm cm}\\
                &+\Big(\frac{d\sigma}{d\Omega}\Big)^{61}_{\gamma p \rightarrow K^0 \Sigma^{+}}\times{N_{\gamma}\epsilon^{60}_{\Sigma^{+}}N_{\rm t}{B_{\Sigma^{+}}}}2\pi \Delta \cos\theta_{\rm cm}.
\end{split}
\end{equation}
Thus, 
\begin{equation}  
\begin{split}
\Big(\frac{d\sigma}{d\Omega}\Big)^{60}_{\gamma n \rightarrow K^0 \Lambda}= \frac{N^{60}_{K^{0}}}{{N_{\gamma}\epsilon^{60}_{\Lambda}N_{\rm t}{B_{\Lambda}}}2\pi \Delta \cos\theta_{\rm cm}} \\
   -~\frac{\epsilon^{60}_{\Sigma^{+}}}{\epsilon^{60}_{\Lambda}}\cdot\frac{B_{\Sigma^{+}}}{B_{\Lambda}}\cdot\Big(\frac{d\sigma}{d\Omega}\Big)^{61}_{\gamma p \rightarrow K^0 \Sigma^{+}}.
\end{split}
\end{equation}
The measured $K^0$ yields for case (60) and the results of a 15-parameter global fit of ${d\sigma}/{d\Omega}$ for $\gamma p \rightarrow K^0 \Sigma^{+}$, discussed in Sec.~\ref{sec:five}.A, were used to calculate the $\gamma n \rightarrow K^0 \Lambda$ cross sections for case (60). It was not possible to determine meaningful values of $(d\sigma/d\Omega)^{60}_{\gamma p \rightarrow K^0 \Sigma^+}$ due to the large subtractions required.

The final task was to determine the $\gamma n \rightarrow K^0 \Sigma^{0}$ cross section for case (70). For this case, recall that $N^{70}_{K^0} =N^{70}_{\Lambda}+N^{70}_{\Sigma^0}$.  Since $(d\sigma/d\Omega)^{70}_{\gamma n \rightarrow K^0 \Lambda}$ = $(d\sigma/d\Omega)^{60}_{\gamma n \rightarrow K^0 \Lambda}$, 
\begin{equation} 
\begin{split}
N^{70}_{K^{0}} &= \Big(\frac{d\sigma}{d\Omega}\Big)^{60}_{\gamma n \rightarrow K^0 \Lambda}\times{N_{\gamma}\epsilon^{70}_{\Lambda}N_{\rm t}{B_{\Lambda}}}2\pi \Delta \cos\theta_{\rm cm}  \\
                &+ \Big(\frac{d\sigma}{d\Omega}\Big)^{70}_{\gamma n \rightarrow K^0 \Sigma^{0}}\times{N_{\gamma}\epsilon^{70}_{\Sigma^{0}}N_{\rm t}{B_{\Sigma^{0}}}}2\pi \Delta \cos\theta_{\rm cm}.
\end{split}
\end{equation}
Thus, 
\begin{equation}  
\begin{split}
\Big(\frac{d\sigma}{d\Omega}\Big)^{70}_{\gamma n \rightarrow K^0 \Sigma^0}= \frac{N^{70}_{K^{0}}}{{N_{\gamma}\epsilon^{70}_{\Sigma^{0}}N_{\rm t}{B_{\Sigma^{0}}}}2\pi \Delta \cos\theta_{\rm cm}}\\
 -~\frac{\epsilon^{70}_{\Lambda}}{\epsilon^{70}_{\Sigma^{0}}}\cdot\frac{B_{\Lambda}}{B_{\Sigma^{0}}}\cdot\Big(\frac{d\sigma}{d\Omega}\Big)^{60}_{\gamma n \rightarrow K^0 \Lambda}.
\end{split}
\end{equation}

The average of the differential cross sections for the cases with and without detection of the final-state neutron, weighted according to the statistical uncertainties, was calculated for  $\gamma n \to K^0 \Lambda$ and $\gamma n \to K^0 \Sigma^0$ and then integrated cross sections were obtained by fitting these values with two-parameter expansions in Legendre polynomials.  The Legendre fits include $P_0$ and $P_1$ terms for the $\gamma n \to K^0 \Lambda$ and $\gamma n \to K^0 \Sigma^0$ results but just a $P_0$ term for the $\gamma p \to K^0 \Sigma^+$ results.  We used only our case (61) results for $\gamma p \to K^0 \Sigma^+$.

\section{\label{sec:four}Calculation of uncertainties}
There are two types of uncertainty involved in calculating the differential cross section. One is the statistical uncertainty and the other is the systematic uncertainty. The statistical uncertainty describes our imprecise knowledge of the kaon signal yield. The systematic uncertainty is the combination of uncertainties from the photon flux, acceptance, and branching ratios. The kaon signal yield in real data was correlated with the centroid of the background. As mentioned earlier, the $\pi^0\pi^0$ invariant-mass distributions were fitted with a sum of scaled Gaussians, with background and signal parts. First the invariant-mass histogram was fitted, the background centroid was noted and the kaon yield was calculated; this is called the nominal case. Next a centroid for the background was chosen, which is an average of nominal case background centroid and kaon signal centroid (498~MeV), and the $m(\pi^0\pi^0)$ distribution was refitted and the kaon yield was recalculated. This is called the modified case. The statistical uncertainty was conservatively calculated using  
\begin{equation}
\label{statuncert}
\Delta N_{K^0} =  [(\rm Poisson\ error)^2+(\rm model\ error)^2]^{\frac{1}{2}}.
\end{equation}
Here, $\rm Poisson\ error$ = $\sqrt{N_{K^0}+1}$, where $N_{K^0}$ is the average number of $K^0$s determined by fitting the $m({\pi^0\pi^0})$ distributions using the nominal and modified values for the background centroid. The model error was taken as the difference in the number of $K^0$s determined using the two different background centroids. The statistical uncertainty in $d\sigma/d\Omega$ is given by
\begin{equation}
\label{statdiffuncer}
 \Delta \Big(\frac{d\sigma}{d\Omega}\Big)_{\rm stat.}= \frac{d\sigma}{d\Omega}\times \Big(\frac{\Delta N_{K^0}}{ N_{K^0}}\Big)
\end{equation}
and the systematic uncertainty is given by 
\begin{equation}
\label{sysdiffuncer}
\Delta \Big(\frac{d\sigma}{d\Omega}\Big)_{\rm sys.}= \frac{d\sigma}{d\Omega}\times 
\Bigg[{\Big(\frac{\Delta N_{\gamma}}{N_{\gamma}}\Big)^2 + 
\Big(\frac{\Delta \epsilon}{\epsilon}\Big)^2 +
\Big(\frac{{\Delta B_{\rm }}}{{B_{\rm }}}\Big)^2}\Bigg]^{\frac{1}{2}},
\end{equation}
where the contribution from the uncertainty in the photon flux varied from 1.1\% to 2.4\% and the contribution from the acceptance varied from about 2\% to about 4\% for $\gamma n \rightarrow K^0 \Lambda$ and
$\gamma n \rightarrow K^0 \Sigma^0$. The contribution from the product of branching ratios was 1.4\% for 
$\gamma n \rightarrow K^0 \Lambda$ and $\gamma n \rightarrow K^0 \Sigma^0$ and was 0.6\% for $\gamma p \rightarrow K^0 \Sigma^+$.
\section{\label{sec:five}Results and discussion}
\subsection{\label{sec:fiveA}$\gamma p\rightarrow K^0\Sigma^+$}
\begin{figure*}[htbp]
	\includegraphics[width=1.0\linewidth]{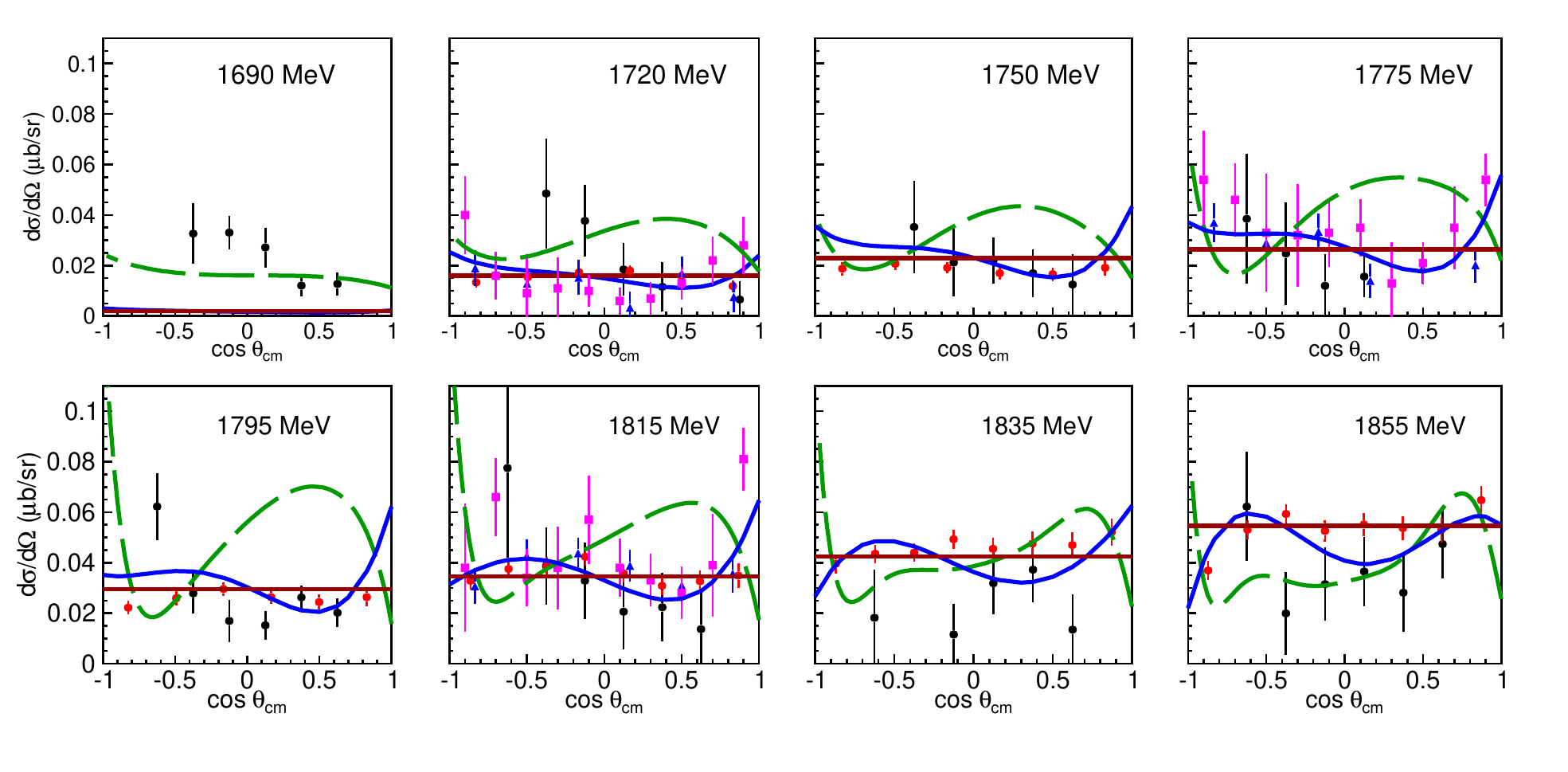}

\caption{Differential cross section for $\gamma p\rightarrow K^0\Sigma^+$ for the various c.m. energy bins. The solid black circles show our results, the solid magenta squares show prior results from Lawall {\it et al.} \cite{lawall}, the solid blue triangles show prior results from Castelijns {\it et al.} \cite{castelijins}, the solid red circles show prior results from Aguar-Bartolom{\'e} {\it et al.} \cite{aguar}, and the dashed green curves represent isobar-model predictions by Mart \cite{tmart}. The solid blue curves show results of a 15-parameter global fit to our results and prior differential cross-section data.  The solid red curves show results of a three-parameter global fit in which the angular distributions were approximated as being isotropic in each energy bin. (See text for details.)}
	\label{K0SpDiffXsec8padsCase61}
\end{figure*}

Figure \ref{K0SpDiffXsec8padsCase61} shows the differential cross section for $\gamma p\rightarrow K^{0}\Sigma^{+}$ for the eight energy bins. 
Our results are shown as solid black circles.  Prior results from Lawall {\it et al.} \cite{lawall}, measured with the SAPHIR detector at ELSA in Bonn, are shown as solid magenta squares. Prior results from Castelijns {\it et al.} \cite{castelijins}, measured with the Crystal Barrel and TAPS spectrometers at ELSA, are shown as solid blue triangles.  The most precise prior results are from Aguar-Bartolom{\'e} {\it et al.} \cite{aguar}, measured on a liquid hydrogen target with the Crystal Ball and TAPS spectrometers at MAMI, and shown as solid red circles.  Our differential cross-section results are in fair agreement within error bars with prior results in the $\cos \theta_{\rm cm}$ range from $+0.6$ to $-0.45$. Our results in the bins at $\cos \theta_{\rm cm} = \pm 0.875$ and $-0.675$ were unreliable, due to the low statistics and low acceptance at these angle bins. Therefore, those results are not shown in Fig.~\ref{K0SpDiffXsec8padsCase61}, nor were they used to calculate the integrated cross sections. The solid blue curves in Fig.~\ref{K0SpDiffXsec8padsCase61} are from a 15-parameter global fit to all the data, which is described below. The solid red curves are from a three-parameter global fit in which the angular distributions were approximated as being isotropic in each energy bin.  The measurements in Fig.~\ref{K0SpDiffXsec8padsCase61} are compared with isobar-model predictions by Mart \cite{tmart}, which are shown as dashed green curves. In general, these predictions do not agree well with the measured angular distributions.

In order to ensure a smooth variation with energy and that the cross section vanishes at threshold, a 15-parameter global fit of our results and prior differential cross-section data was performed.  This fit used the parametrization
\begin{equation}
\frac{d\sigma}{d\Omega} = \sum_{n=1}^{3} \sum_{\ell=0}^4 a_{n\ell} (W - W_T)^n  P_\ell(\cos \theta_{\rm cm}),
\end{equation}
where $W_T = 1687$~MeV is the threshold energy for $\gamma p \to K^0 \Sigma^+$ and $P_\ell(\cos \theta_{\rm cm})$ is a Legendre polynomial.  The $a_{n\ell}$ coefficients were constant fitting parameters.  Uncertainties in the fitted cross sections were conservatively calculated as twice the difference between results of the 15-parameter global fit and a separate three-parameter global fit in which the angular distributions were approximated as being isotropic in each energy bin (only the $a_{n0}$ coefficients were varied). 

\begin{figure}
\begin{center}
	\includegraphics[scale=0.4]{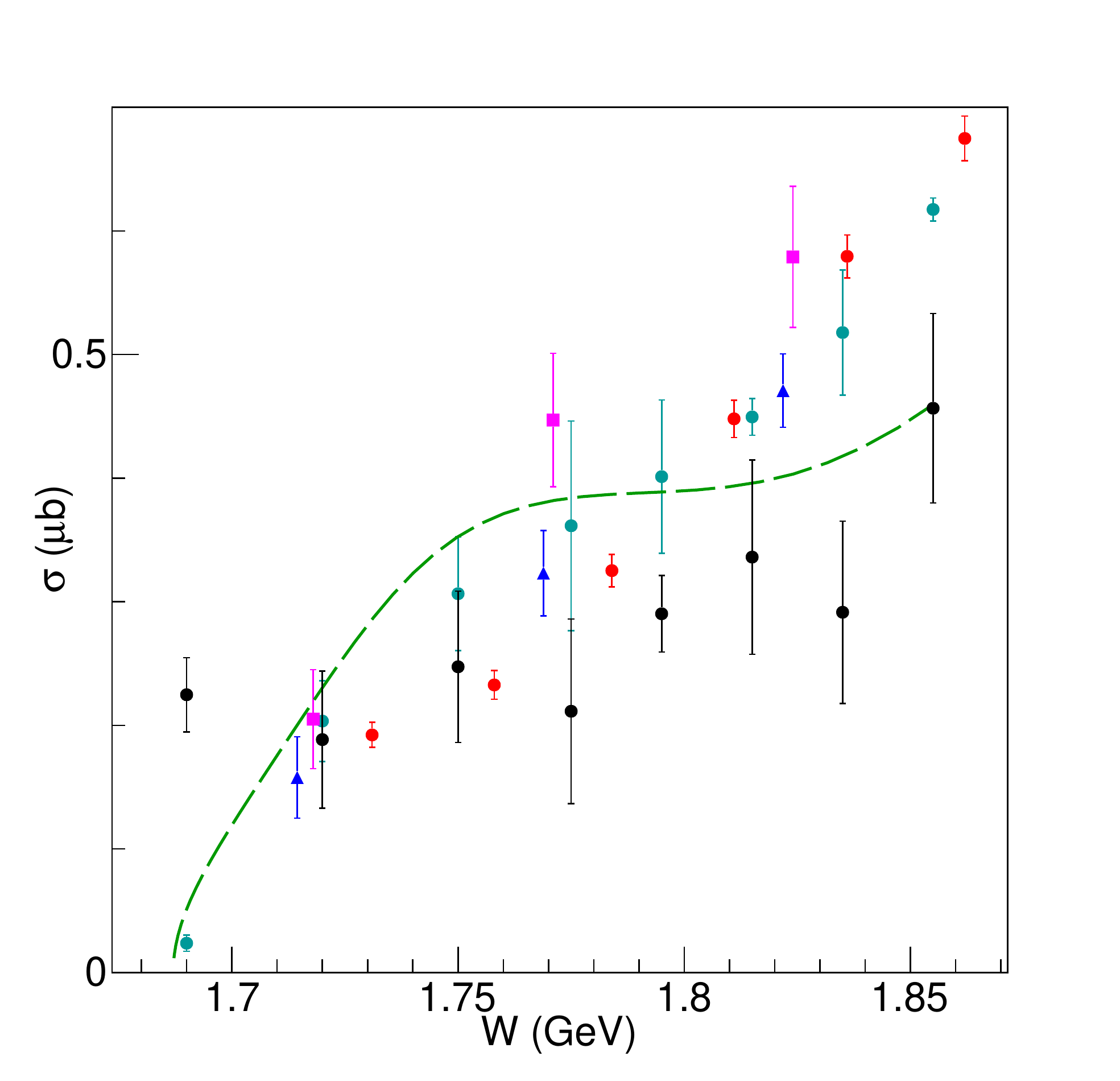}
\caption{ Integrated cross section for $\gamma p\rightarrow K^0\Sigma^+$. Our results, shown as solid black circles, were obtained by fitting our measured differential cross sections with a one-parameter Legendre expansion.  The solid magenta squares show prior results from Lawall {\it et al.} \cite{lawall}, the solid blue triangles show prior results from Castelijns {\it et al.} \cite{castelijins}, the solid red circles show prior results from Aguar-Bartolom{\'e} {\it et al.} \cite{aguar}, and the dashed green curve represents an isobar-model prediction by Mart \cite{tmart}. The solid cyan circles were obtained from a 15-parameter global fit of our results combined with prior differential cross-section data.  (See text for details.)}
	\label{k0sptotalX}
	\end{center}
\end{figure}

Our measured integrated cross sections for $\gamma p\rightarrow K^{0}\Sigma^{+}$ were obtained by making one-parameter Legendre fits of our measured differential cross sections.  They are shown in Fig.~\ref{k0sptotalX} as solid black circles. 
Prior results from Lawall {\it et al.}  \cite{lawall}, Castelijns {\it et al.} \cite{castelijins}, and Aguar-Bartolom{\'e} {\it et al.} \cite{aguar}
are shown as solid magenta squares, solid blue triangles, and solid red circles, respectively.  The results of our 15-parameter global fit are shown as solid cyan circles. The experimental results are compared with Mart's isobar-model predictions \cite{tmart} shown as a dashed green curve. 

\subsection{\label{sec:fiveB}$\gamma n \rightarrow K^0 \Lambda$}
\begin{figure*}[htbp]
\centering

		\includegraphics[width=1.0\linewidth]{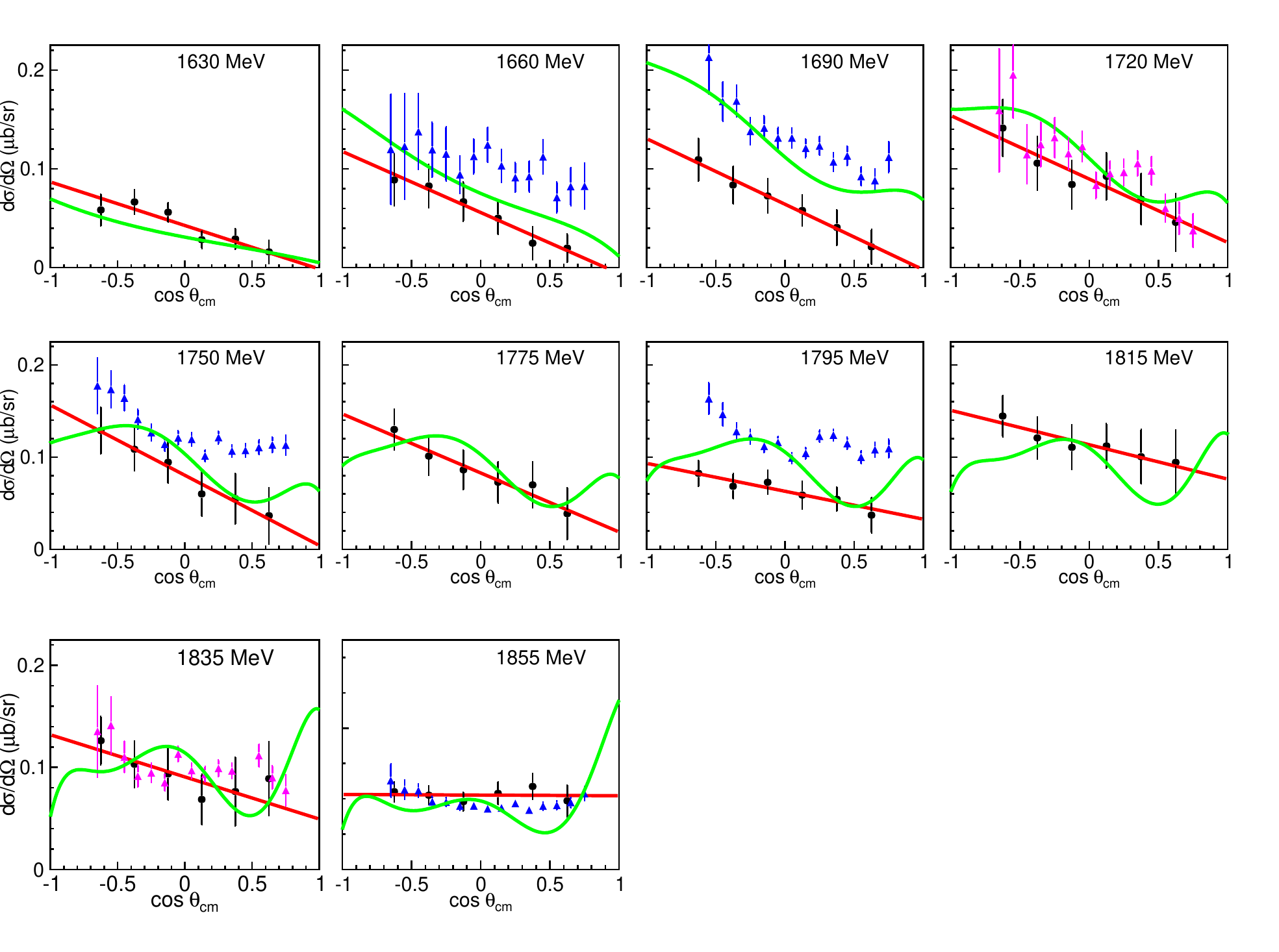}
	
\caption{Differential cross section for $\gamma n\rightarrow K^0\Lambda$. The solid black circles represent the weighted average of our results for cases (60) and (70). The solid magenta triangles and solid blue triangles respectively show g10 and g13 results from Compton {\it et al.} \cite{nickcompton}. The solid green curves are a prediction \cite{brian} based upon a partial-wave analysis.  The solid red curves show the results of two-parameter Legendre polynomial fits to our measurements.}
	\label{K0L0DiffXsec10padsCase6070}
\end{figure*}

Since the measured $\gamma p \rightarrow K^0 \Sigma^{+}$ cross sections for case (61) were imprecise due to low statistics and the low acceptance at backward and forward angles, the fitted world values of 
$(d\sigma/d\Omega)_{\gamma p \rightarrow K^0 \Sigma^{+}}$ and the measured $K^0$ yields for case (60) were used to calculate $\gamma n \rightarrow K^0 \Lambda$ cross sections for case (60). Because the associated uncertainties in the fitted world values were relatively large at $\cos\theta_{\rm cm} = \pm 0.875$, those angle bins were excluded for all three $K^0$ photoproduction reactions.  The c.m.\ energy range $W=1615$ to $1765$~MeV was divided into five bins of width 30~MeV, and the c.m.~energy range $W=1765 ~\rm to ~1865$~MeV was divided into five bins of width 20~MeV. The first two c.m.\ energy bins $W=1630$ and $1660$~MeV were below $\gamma p \rightarrow K^0\Sigma^{+}$ threshold 1687~MeV. Therefore only $\gamma n \rightarrow K^0 \Lambda$ events can contribute to these bins. Figure \ref{K0L0DiffXsec10padsCase6070} shows the differential cross section for $\gamma n\rightarrow K^{0}\Lambda$ for these ten energy bins. 
Solid black circles show our results (weighted average of cases (60) and (70)). The solid magenta triangles and solid blue triangles respectively show the g10 and g13 results from Compton {\it et al.} \cite{nickcompton} measured at JLab.  Our results agree, within uncertainties, with the JLab g10 results in the energy bins at 1720 and 1835~MeV and with the JLab g13 results in the energy bin at 1855~MeV. Our results are similar in shape to the JLab g13 measurements at 1660, 1690, 1750, and 1795~MeV but are smaller in magnitude. It should be noted that the g10 and g13 results, where they overlap, are consistent for c.m.\ energies above about 1800~MeV, but the g13 results below that energy are {\it all} larger (especially at forward angles) than the g10 result that falls into our energy bin at 1690~MeV.  The solid red curves in Fig.~\ref{K0L0DiffXsec10padsCase6070} show results of two-parameter Legendre polynomial fits to our measurements.  The solid green curves show predictions based upon a partial-wave analysis \cite{brian}. Our results are in fair agreement with the predictions in all energy bins except at 1690~MeV.

\begin{figure*}[htbp]
\centering
		\includegraphics[width=1.0\linewidth]{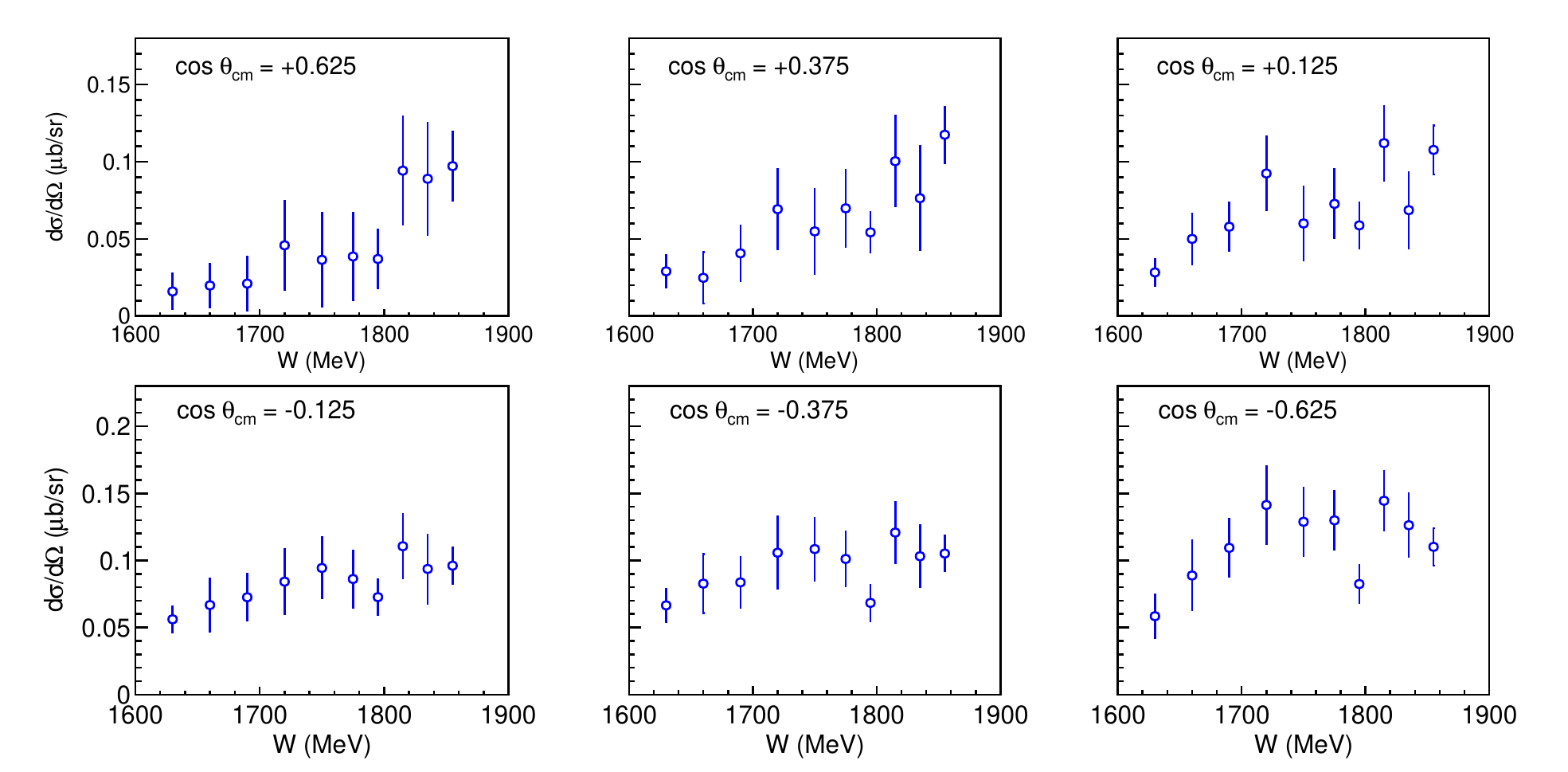}
	
\caption{Differential cross section for $\gamma n\rightarrow K^0\Lambda$ versus c.m.~energy $W$ for angle bins from $\cos \rm{\theta_{cm}}=+0.625$ to $-0.625$. The open blue circles represent the weighted average of our results for cases (60) and (70).}
	\label{K0L0DiffVsTheta}
\end{figure*}

We have checked various factors that might affect the normalizations of our results ({\it e.g.}, the photon flux $N_\gamma$ and detector acceptance) and have been unable to find any problems that would explain the differences between our results and the low-energy g13 results.  Our results for all energy bins were handled in exactly the same manner as each other.  Figure~\ref{K0L0DiffVsTheta} shows the differential cross section for $\gamma n \rightarrow K^0 \Lambda$ as a function of c.m.\ energy $W$ for individual angle bins.  The results in this plot show a generally smooth energy variation, which implies we do not have normalization inconsistencies in individual energy bins.

\begin{figure}
\begin{center}
	\includegraphics[scale=0.4]{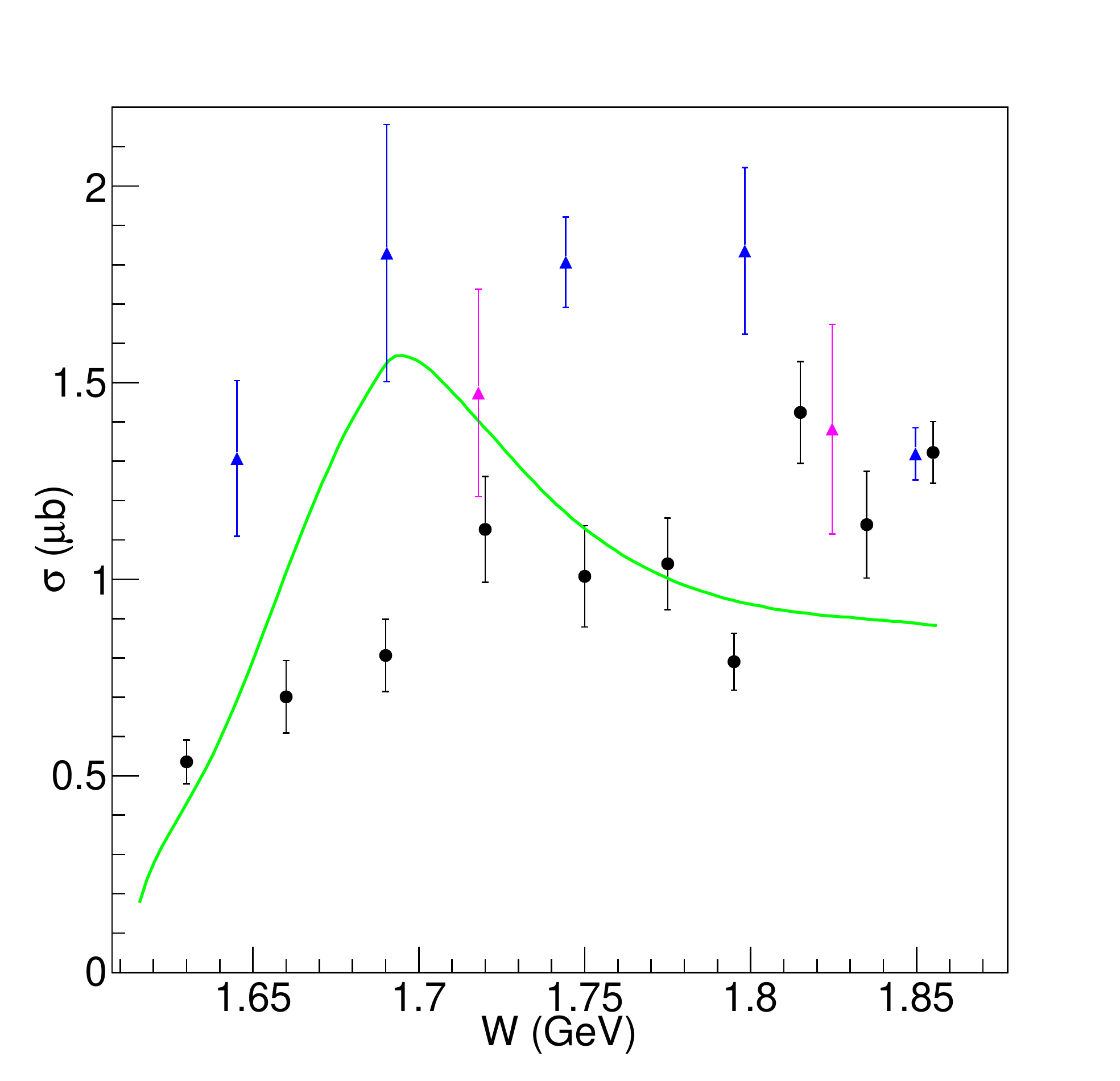}
\caption{ Integrated cross section for $\gamma n\rightarrow K^0\Lambda$. The solid black circles represent our results.  The solid magenta triangles and solid blue triangles respectively show the g10 and g13 results from Compton {\it et al.} \cite{nickcompton}.  The solid green curve shows a prediction \cite{brian} based upon a partial-wave analysis.}
	\label{KL6070AveTotalCrossection}
	\end{center}
\end{figure}

Measured integrated cross sections for $\gamma n\rightarrow K^{0}\Lambda$ are shown in Fig.~\ref{KL6070AveTotalCrossection}. Solid black circles show our results, which were obtained by making two-parameter Legendre fits of the weighted average of our measured $\gamma n\rightarrow K^{0}\Lambda$ differential cross sections for cases (60) and (70).  The solid magenta triangles and solid blue triangles respectively show the g10 and g13 results from Compton {\it et al.} \cite{nickcompton} measured at JLab. The solid green curve shows a prediction based upon a partial-wave analysis \cite{brian}.

\subsection{\label{sec:fiveC}$\gamma n\rightarrow K^{0}\Sigma^0$}
\begin{figure*}[htbp]

\centering
		\includegraphics[width=1.0\linewidth]{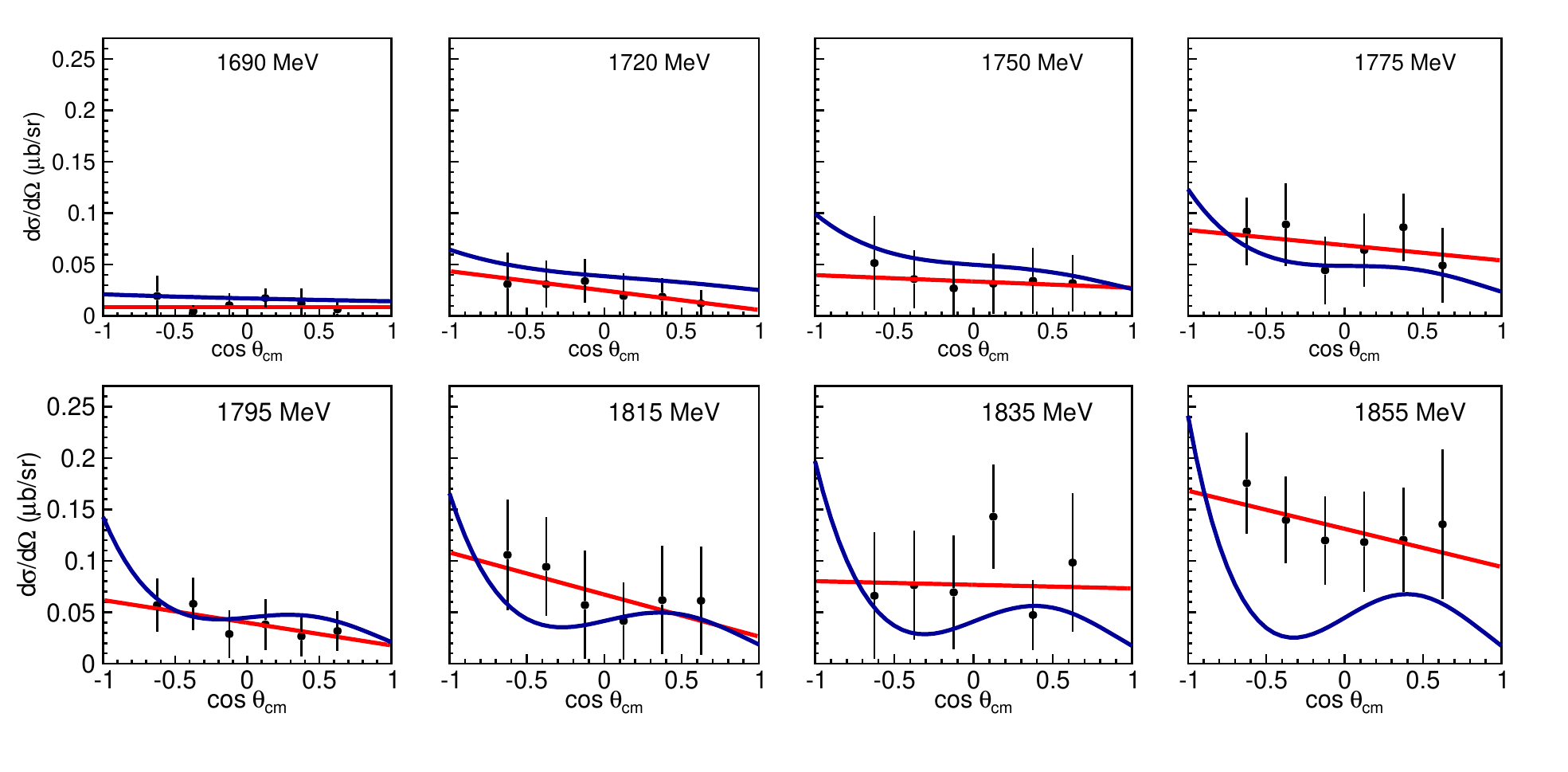}

\caption{Differential cross section for $\gamma n\rightarrow K^0\Sigma^0$. Solid black circles show our results. Solid blue curves represent isobar-model predictions by Mart \cite{tmart} and the solid red curves show results of two-parameter Legendre polynomial fits to our measurements.}
	\label{K0S0DiffXsec8pads}
\end{figure*}

\begin{figure*}[htbp]
\centering
		\includegraphics[width=1.0\linewidth]{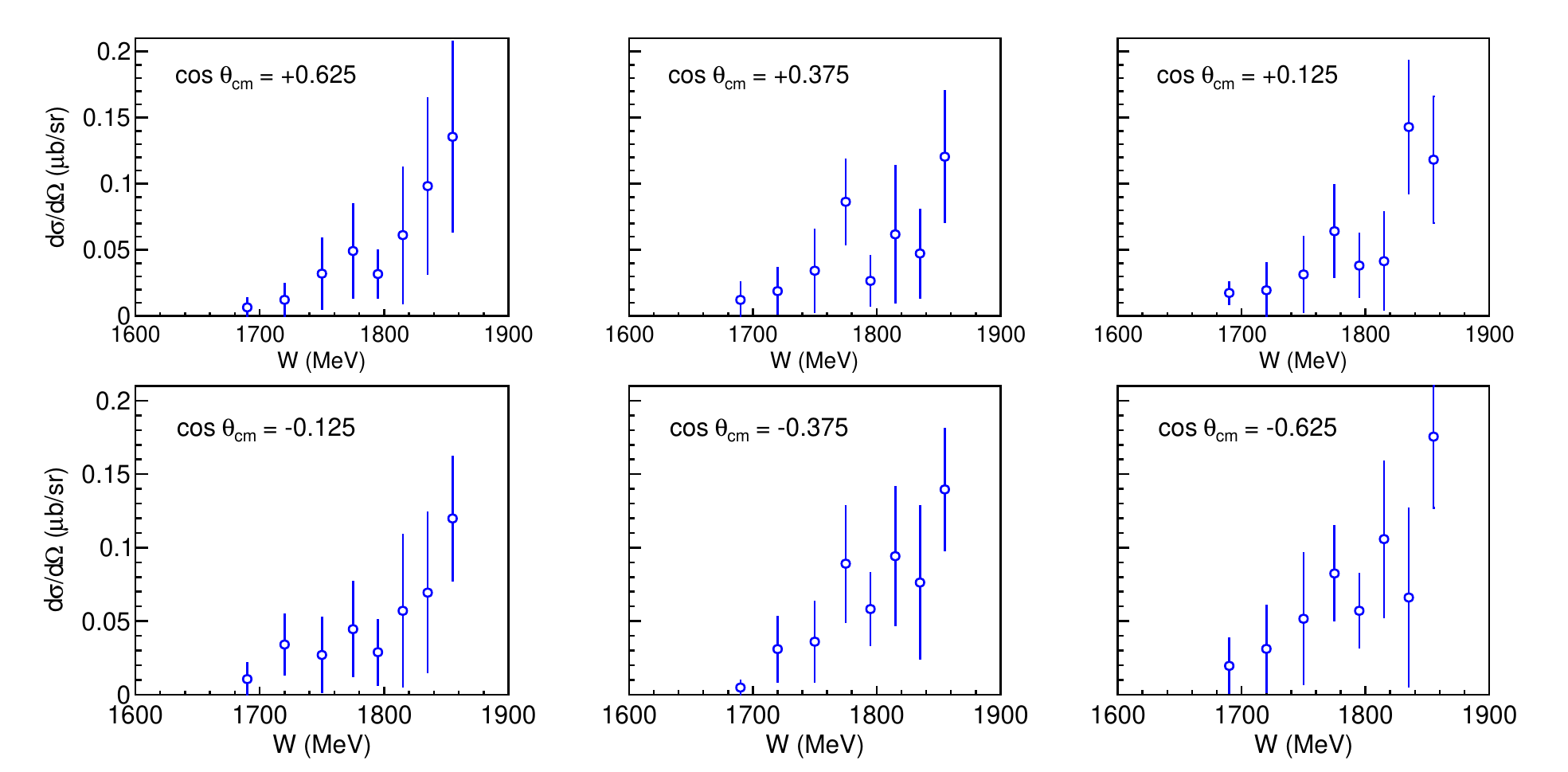}
	
\caption{Differential cross section for $\gamma n\rightarrow K^0\Sigma^{0}$ versus c.m.~energy $W$ for angle bins from $\cos \rm{\theta_{cm}}=+0.625$ to $-0.625$. The open blue circles represent the weighted average of our results for cases (70) and (80).}
	\label{K0S0DiffVsTheta}
\end{figure*}

Our measured $\gamma n \rightarrow K^0 \Lambda$ differential cross sections for case (60) and our measured $K^0$ yields for case (70) were used to calculate the $\gamma n \rightarrow K^0 \Sigma^0$ differential cross sections for case  (70). The c.m.\ energy range $W=1675$ to $1765$~MeV was divided into three bins of width 30~MeV, and the c.m.~energy range $W=1765$ to $1865$~MeV was divided into five bins of width 20~MeV. Figure \ref{K0S0DiffXsec8pads} shows the differential cross section for $\gamma n\rightarrow K^{0}\Sigma^0$ (weighted average of cases (70) and (80)) for these eight c.m.\ energy bins. Our results are compared with isobar-model predictions (dashed blue curves) by Mart \cite{tmart} and the solid red curves show results of two-parameter Legendre polynomial fits to our measurements. Our differential cross section results are in good agreement within error bars with Mart's predictions except at the highest energy bin, $W=1855$~MeV.  Figure~\ref{K0S0DiffVsTheta} shows the differential cross section for $\gamma n \rightarrow K^0 \Sigma^0$ as a function of c.m.\ energy $W$ for individual angle bins.  As for $\gamma n \rightarrow K^0 \Lambda$, these results show a generally smooth energy variation, which supports the fact that the normalizations were determined consistently for the different energy bins.

\begin{figure}
\begin{center}
		\includegraphics[scale=0.4]{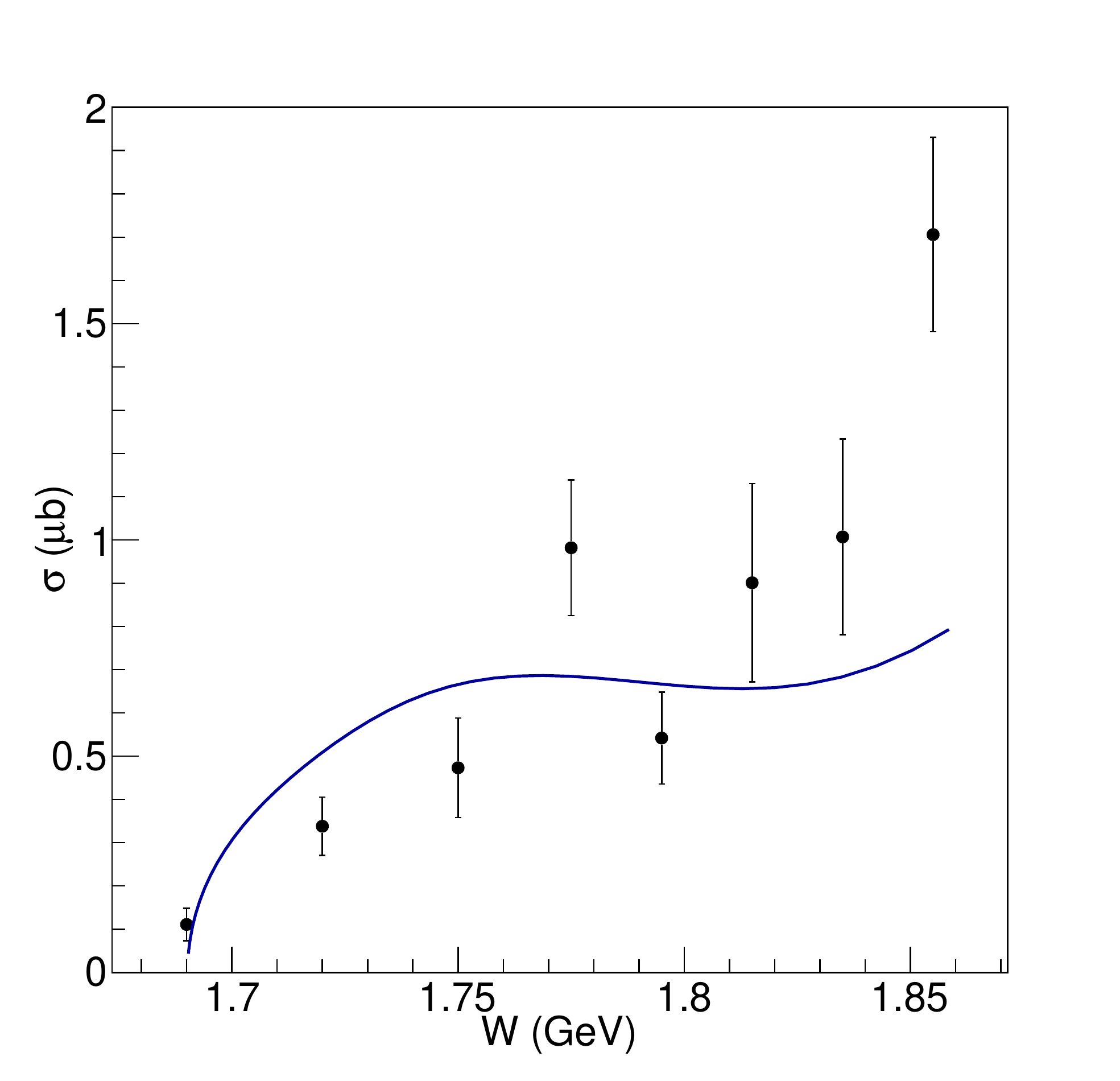}

\caption{Integrated cross section for $\gamma n\rightarrow K^0\Sigma^0$. Solid black circles show our results. The solid blue curve represents an isobar-model prediction by Mart \cite{tmart}.}
	\label{KS7080AveTotalCrossection}
	\end{center}
\end{figure}

Our measured integrated cross section values for $\gamma n\rightarrow K^{0}\Sigma^0$ are shown in Fig.~\ref {KS7080AveTotalCrossection} as solid black circles.   Our integrated cross sections were obtained by calculating the weighted average of our differential cross sections for cases (70) and (80) and then making two-parameter Legendre fits.  Our experimental results are compared with an isobar-model prediction (solid blue curve) by Mart \cite{tmart}. Our results are in good agreement with Mart's predictions except at the highest energy. These are the first experimental results for $\gamma n\rightarrow K^{0}\Sigma^0$.  As in the case of the differential cross sections, our results are in good agreement with Mart's predictions except at the highest energy bin.

\section{\label{sec:six}Summary and Conclusions}

Our results for $\gamma p\rightarrow K^{0}\Sigma^{+}$ are in fair agreement with prior measurements in the $\cos \theta_{\rm cm}$ range from $-$0.45 to $+$0.6, but our results at the most forward and backward angles are unreliable. For this reason, we used $\gamma p\rightarrow K^{0}\Sigma^{+}$ world data to extract the $\gamma n \rightarrow K^0 \Lambda$ cross section for case (60). An isobar-model prediction by Mart \cite{tmart} generally disagrees with all the measured differential cross sections.

Only one published set of prior measurements for $\gamma n\rightarrow K^0\Lambda$ was available for comparing with our results. These prior results were measured with the CLAS spectrometer at JLab \cite{nickcompton} in two separate datasets. In the seven energy bins where our results can be compared, our results agree within uncertainties with the g10 results but our results have a somewhat similar shape, but smaller magnitude, compared with the g13 results below $W = 1800$~MeV.  The results presented in Ref.~\cite{nickcompton} show that the g10 and g13 results, where they overlap, are generally consistent above about $W = 1800$~MeV but {\it not} at lower energies.  Our results are in fairly good agreement (except at 1690~MeV) with a prediction based on a partial-wave analysis \cite{brian} within error bars. Our results for $\gamma n \rightarrow K^0 \Lambda$ provide new measurements in the c.m.~energy range from threshold (1614~MeV) to 1855~MeV. 

Our results for $\gamma n\rightarrow K^{0}\Sigma^{0}$ are the first experimental results for that reaction and span the c.m.~energy range from the threshold (1691~MeV) to 1855~ MeV. Our differential cross sections for $\gamma n\rightarrow K^{0}\Sigma^{0}$ are in good agreement within error bars with isobar-model predictions by Mart \cite{tmart} except at the highest energy bin. Our two independent measurements for cases (70) and (80) are consistent within error bars. 

In summary, our new cross-section measurements for $\gamma n \rightarrow K^0 \Lambda$ and $\gamma n \rightarrow K^0 \Sigma^0$ will provide valuable data for future partial-wave analyses and will help better determine the properties of $N^*$ resonances that decay to $K\Lambda$ or $K\Sigma$ final states.

\begin{acknowledgments}
The authors would like to thank Drs. Brian Hunt and Terry Mart for providing the model predictions for $\gamma n \rightarrow K^0 \Lambda$ and $\gamma n \rightarrow K^0 \Sigma^0$ reactions, respectively. This work was supported in part by the U.S.\ Department of Energy, Office of Science, Office of Nuclear Physics, under Awards No.\ DE-FG02-01ER41194 and DE-SC0014323, and by the Department of Physics at Kent State University. We would like to thank all the technical and non-technical staff of MAMI for their support. This work was supported by Schweizerischer Nationalfonds (200020- 132799, 121781, 117601, 113511), Deutsche Forschungsgemeinschaft
(SFB 443, SFB/TR 16), DFG-RFBR (Grant No.05-02-04014), UK Science and Technology Facilities Council(STFC 57071/1, 50727/1), European Community Research Infrastructure Activity (FP6), the U.S. DOE, U.S. NSF, and NSERC (Canada).
\end{acknowledgments}





\appendix*
\section{Tabulation of Results}
In this appendix, we provide our measured differential and integrated cross sections for $\gamma p \rightarrow K^0 \Sigma^+$ in Tables \ref{A} and \ref{B}, our measured differential and integrated cross sections for $\gamma n \rightarrow K^0 \Lambda$ in Tables \ref{C} and \ref{D}, and our measured differential and integrated cross sections for $\gamma n \rightarrow K^0 \Sigma^0$ in Tables \ref{E} and \ref{F}.

\begin{table*}[htb]
\caption{Differential cross section for $\gamma p \rightarrow K^0 \Sigma^+$. Systematic uncertainties less than 0.001 are not listed.}
\begin{ruledtabular}
\begin{tabular}{cccccccccc}
$W$ &$\cos{\theta_{\rm cm}}$ & $d\sigma/d\Omega$ & stat.\ unc.\  & sys.\ unc.& $W$ &$\cos{\theta_{\rm cm}}$ & $d\sigma/d\Omega$ & stat.\ unc.\  & sys.\ unc.\\
(MeV)&& ($\mu$b/sr) & ($\mu$b/sr) & ($\mu$b/sr)&(MeV)&& ($\mu$b/sr) & ($\mu$b/sr) & ($\mu$b/sr)\\
\hline

${1690}	$&${+0.625}	$&$0.013	$&$0.004	$&$0.001        $&${1795}	$&${+0.625}	$&$	0.020$&$	0.005$&$  0.001		$ \\  
${1690}	$&${+0.375}	$&$0.012	$&$0.004	$&$0.001	$&${1795}	$&${+0.375}	$&$	0.026$&$	0.004$&$      0.001		$ \\ 
${1690}	$&${+0.125}	$&$0.027	$&$0.007	$&$0.002	$&${1795}	$&${+0.125}	$&$	0.015$&$	0.005$&$      0.001		$ \\ 
${1690}	$&${-0.125}	$&$0.033	$&$0.006	$&$0.002	$&${1795}	$&${-0.125}	$&$	0.017$&$	0.008$&$      0.001		$  \\ 
${1690}	$&${-0.375}	$&$0.033	$&$0.012	$&$0.002	$&${1795}	$&${-0.375}	$&$	0.028$&$	0.007$&$      0.001		$ \\ 
${1720}	$&${+0.625}     $&$	-       $&$	-       $&$	-	$&${1815}	$&${+0.625}     $&$	0.014$&$	0.014$&$      0.001		$\\
${1720}	$&${+0.375}     $&$0.012       $&$0.009       $&$0.001	$&${1815}	$&${+0.375}     $&$	0.022$&$	0.013$&$      0.002		$ \\ 
${1720}	$&${+0.125}     $&$0.018        $&$0.010        $&$0.001	$&${1815}	$&${+0.125}     $&$	0.021$&$	0.015$&$      0.001		$ \\ 
${1720}	$&${-0.125}     $&$0.038        $&$0.014        $&$0.002	$&${1815}	$&${-0.125}     $&$	0.033$&$	0.015$&$      0.002		$ \\ 
${1720}	$&${-0.375}     $&$0.049        $&$0.021        $&$0.002	$&${1815}	$&${-0.375}     $&$	0.039$&$	0.015$&$      0.003		$ \\ 
${1750}	$&${+0.625}	$&$0.012       $&$0.012       $&$-	$&${1835}	$&${+0.625}     $&$	0.013$&$	0.014$&$	0.001	$ \\ 
${1750}	$&${+0.375}	$&$0.017       $&$0.009       $&$0.001	$&${1835}	$&${+0.375}     $&$	0.037$&$	0.012$&$	0.003	$  \\ 
${1750}	$&${+0.125}	$&$0.022       $&$0.009       $&$0.001	$&${1835}	$&${+0.125}     $&$	0.032$&$	0.012$&$	0.003	$  \\ 
${1750}	$&${-0.125}	$&$0.021       $&$0.013       $&$0.001	$&${1835}	$&${-0.125}     $&$	0.012$&$	0.012$&$	0.001	$  \\
${1750}	$&${-0.375}	$&$0.035        $&$0.018        $&$0.001	$&${1835}	$&${-0.375}     $&$	0.003$&$	0.003$&$	-	$  \\
${1775}	$&${+0.625}     $&$0.005       $&$0.005       $&$-	$&${1855}	$&${+0.625}     $&$	0.047$&$	0.013$&$      0.004		$\\ 
${1775}	$&${+0.375}     $&$0.006       $&$0.006       $&$-	$&${1855}	$&${+0.375}     $&$	0.028$&$	0.015$&$      0.002		$  \\ 
${1775}	$&${+0.125}     $&$0.016       $&$0.008       $&$0.001	$&${1855}	$&${+0.125}     $&$	0.037$&$	0.013$&$      0.003		$\\ 
${1775}	$&${-0.125}     $&$0.012       $&$0.012       $&$0.001	$&${1855}	$&${-0.125}     $&$	0.032$&$	0.014$&$      0.003		$\\ 
${1775}	$&${-0.375}     $&$0.025       $&$0.020       $&$0.001	$&${1855}	$&${-0.375}     $&$	0.020$&$	0.016$&$      ~0.002		$  
	\label{A}
\end{tabular}
\end{ruledtabular}
\end{table*}

\begin{table}[htb]
\caption{Integrated cross section for $\gamma p \rightarrow K^0 \Sigma^+$.}
\begin{ruledtabular}
\begin{tabular}{cccc}

$W$ & ${\sigma }$& stat.\ unc. & sys.\ unc.\\
(\rm MeV)   & $(\mu b)$& ($\mu$b) & ($\mu$b) \\
\hline
$1690$	&    ${0.225}$&$0.030$&$0.010 $ \\ 
$1720$	&    ${0.188}$&$0.055$&$0.006$ \\  
$1750$	&    ${0.247}$&$0.061$&$0.005$ \\  
$1775$	&    ${0.211}$&$0.075$&$0.007$ \\  
$1795$	&    ${0.290}$&$0.031$&$0.011$ \\  
$1815$	&    ${0.336}$&$0.079$&$0.023$ \\  
$1835$	&    ${0.291}$&$0.074$&$0.023$ \\  
$1855$	&    ${0.457}$&$0.077$&$~0.035$
	\label{B}
\end{tabular}
\end{ruledtabular}
\end{table}

\begin{table*}[htb]
\caption{Differential cross section for $\gamma n \rightarrow K^0 \Lambda$.}
\begin{ruledtabular}
\begin{tabular}{cccccccccc}

$W$ &$\cos{\theta_{\rm cm}}$ & $d\sigma/d\Omega$ & stat.\ unc.\  &
sys.\ unc.& $W$ &$\cos{\theta_{\rm cm}}$ & $d\sigma/d\Omega$ & stat.\
unc.\  & sys.\ unc.\\
(MeV)&& ($\mu$b/sr) & ($\mu$b/sr) & ($\mu$b/sr)&(MeV)&& ($\mu$b/sr) &
($\mu$b/sr) & ($\mu$b/sr)\\
\hline
${1630}$&	${+0.625}$&	${0.016}$&	${0.011}$&	${0.001}$&${1775}$&	${+0.625}$&	${0.039}$&	${0.028}$&	$0.002$ \\
${1630}$&	${+0.375}$&	${0.029}$&	${0.010}$&	${0.001}$&${1775}$&	${+0.375}$&	${0.070}$&	${0.025}$&	$0.003$ \\
${1630}$&	${+0.125}$&	${0.028}$&	${0.009}$&	${0.001}$&${1775}$&	${+0.125}$&	${0.073}$&	${0.022}$&	${0.003}$ \\
${1630}$&	${-0.125}$&	${0.056}$&	${0.010}$&	${0.002}$&${1775}$&	${-0.125}$&	${0.086}$&	${0.021}$&	${0.004}$ \\
${1630}$&	${-0.375}$&	${0.066}$&	${0.012}$&	${0.003}$&${1775}$&	${-0.375}$&	${0.101}$&	${0.020}$&	${0.004}$ \\
${1630}$&	${-0.625}$&	${0.058}$&	${0.016}$&	${0.002}$&${1775}$&	${-0.625}$&	${0.130}$&	${0.022}$&	$0.005$\\  \hline
${1660}$&	${+0.625}$&	${0.020}$&	${0.014}$&	${0.001}$&${1795}$&	${+0.625}$&	${0.037}$&	${0.019}$&	$0.002$\\
${1660}$&	${+0.375}$&	${0.025}$&	${0.017}$&	${0.001}$&${1795}$&	${+0.375}$&	${0.054}$&	${0.013}$&	$0.002$\\
${1660}$&	${+0.125}$&	${0.050}$&	${0.016}$&	${0.002}$&${1795}$&	${+0.125}$&	${0.059}$&	${0.015}$&	$0.003$\\
${1660}$&	${-0.125}$&	${0.067}$&	${0.020}$&	${0.003}$&${1795}$&	${-0.125}$&	${0.073}$&	${0.013}$&	$0.003$\\
${1660}$&	${-0.375}$&	${0.083}$&	${0.022}$&	${0.004}$&${1795}$&	${-0.375}$&	${0.068}$&	${0.013}$&	$0.003$\\
${1660}$&	${-0.625}$&	${0.089}$&	${0.026}$&	${0.004}$&${1795}$&	${-0.625}$&	${0.082}$&	${0.014}$&	$0.004$\\  \hline
${1690}$&	${+0.625}$&	${0.021}$&	${0.017}$&	${0.001}$&${1815}$&	${+0.625}$&	${0.094}$&	${0.035}$&	${0.007}$\\
${1690}$&	${+0.375}$&	${0.041}$&	${0.018}$&	${0.002}$&${1815}$&	${+0.375}$&	${0.100}$&	${0.029}$&	${0.007}$\\
${1690}$&	${+0.125}$&	${0.058}$&	${0.016}$&	${0.003}$&${1815}$&	${+0.125}$&	${0.112}$&	${0.024}$&	${0.008}$\\
${1690}$&	${-0.125}$&	${0.073}$&	${0.017}$&	${0.004}$&${1815}$&	${-0.125}$&	${0.111}$&	${0.024}$&	${0.008}$\\
${1690}$&	${-0.375}$&	${0.084}$&	${0.019}$&	${0.004}$&${1815}$&	${-0.375}$&	${0.121}$&	${0.023}$&	${0.009}$\\
${1690}$&	${-0.625}$&	${0.109}$&	${0.022}$&	${0.005}$&${1815}$&	${-0.625}$&	${0.144}$&	${0.022}$&	${0.010}$\\  \hline
${1720}$&	${+0.625}$&	${0.046}$&	${0.029}$&	${0.002}$&${1835}$&	${+0.625}$&	${0.089}$&	${0.036}$&	${0.007}$\\
${1720}$&	${+0.375}$&	${0.069}$&	${0.026}$&	${0.003}$&${1835}$&	${+0.375}$&	${0.076}$&	${0.034}$&	${0.006}$\\
${1720}$&	${+0.125}$&	${0.092}$&	${0.024}$&	${0.004}$&${1835}$&	${+0.125}$&	${0.069}$&	${0.024}$&	${0.006}$\\
${1720}$&	${-0.125}$&	${0.084}$&	${0.024}$&	${0.003}$&${1835}$&	${-0.125}$&	${0.094}$&	${0.026}$&	${0.008}$\\
${1720}$&	${-0.375}$&	${0.106}$&	${0.027}$&	${0.004}$&${1835}$&	${-0.375}$&	${0.103}$&	${0.023}$&	${0.008}$\\
${1720}$&	${-0.625}$&	${0.141}$&	${0.029}$&	${0.005}$&${1835}$&	${-0.625}$&	${0.126}$&	${0.023}$&	${0.010}$\\ \hline
${1750}$&	${+0.625}$&	${0.036}$&	${0.030}$&	${0.001}$&${1855}$&	${+0.625}$&	${0.097}$&	${0.022}$&	${0.008}$\\
${1750}$&	${+0.375}$&	${0.055}$&	${0.027}$&	${0.002}$&${1855}$&	${+0.375}$&	${0.118}$&	${0.018}$&	${0.010}$\\
${1750}$&	${+0.125}$&	${0.060}$&	${0.024}$&	${0.002}$&${1855}$&	${+0.125}$&	${0.108}$&	${0.016}$&	${0.009}$\\
${1750}$&	${-0.125}$&	${0.094}$&	${0.023}$&	${0.003}$&${1855}$&	${-0.125}$&	${0.096}$&	${0.013}$&	${0.008}$\\
${1750}$&	${-0.375}$&	${0.108}$&	${0.023}$&	${0.003}$&${1855}$&	${-0.375}$&	${0.105}$&	${0.013}$&	${0.008}$\\
${1750}$&	${-0.625}$&	${0.129}$&	${0.025}$&	${0.003}$&${1855}$&	${-0.625}$&	${0.110}$&	${0.014}$&	$~{0.009}$
        \label{C}
\end{tabular}
\end{ruledtabular}
\end{table*}

\begin{table}[htb]
\caption{Integrated cross section for $\gamma n \rightarrow K^0 \Lambda$.}
\begin{ruledtabular}
\begin{tabular}{cccc}
$W$ & ${\sigma }$& stat.\ unc. & sys.\ unc.\\
(\rm MeV)   & $(\mu b)$& ($\mu$b) & ($\mu$b) \\
\hline
${1630}$&	${0.54}$&	${0.05}$&	${0.02}$\\
${1660}$&	${0.70}$&	${0.09}$&	${0.03}$\\
${1690}$&	${0.81}$&	${0.09}$&	${0.03}$\\
${1720}$&	${1.13}$&	${0.13}$&	${0.02}$\\
${1750}$&	${1.01}$&	${0.13}$&	${0.04}$\\
${1775}$&	${1.04}$&	${0.12}$&	${0.04}$\\
${1795}$&	${0.79}$&	${0.07}$&	${0.05}$\\
${1815}$&	${1.42}$&	${0.13}$&	${0.11}$\\
${1835}$&	${1.14}$&	${0.13}$&	${0.09}$\\
${1855}$&	${1.32}$&	${0.08}$&	$~{0.02}$
        \label{D}
\end{tabular}
\end{ruledtabular}
\end{table}

\begin{table*}[htb]
\caption{Differential cross section for $\gamma n \rightarrow K^0 \Sigma^0$. Systematic uncertainties less than 0.001 are not listed.}
\begin{ruledtabular}
\begin{tabular}{cccccccccc}
$W$ &$\cos{\theta_{\rm cm}}$ & $d\sigma/d\Omega$ & stat.\ unc.\  & sys.\ unc.& $W$ &$\cos{\theta_{\rm cm}}$ & $d\sigma/d\Omega$ & stat.\ unc.\  & sys.\ unc.\\
(MeV)&& ($\mu$b/sr) & ($\mu$b/sr) & ($\mu$b/sr)&(MeV)&& ($\mu$b/sr) & ($\mu$b/sr) & ($\mu$b/sr)\\
\hline

${1690}	$&${+0.625}	$&$0.007	$&$0.007	$&$ - $& ${1795}	$&${+0.625}	$&$0.032	$&$0.018	$&$ 0.002 $ \\ 	
${1690}	$&${+0.375}	$&$0.012	$&$0.013	$&$ 0.001 $& ${1795}	$&${+0.375}	$&$0.027	$&$0.019	$&$ 0.001 $ \\ 
${1690}	$&${+0.125}	$&$0.017	$&$0.008	$&$ 0.001 $& ${1795}	$&${+0.125}	$&$0.038	$&$0.024	$&$ 0.002 $ \\ 
${1690}	$&${-0.125}	$&$0.011	$&$0.011	$&$ 0.001 $&${1795}	$&${-0.125}	$&$0.029	$&$0.022	$&$ 0.001 $  \\ 
${1690}	$&${-0.375}	$&$0.005	$&$0.005	$&$ - $&${1795}	$&${-0.375}	$&$0.058	$&$0.025	$&$ 0.003$   \\ 
${1690}	$&${-0.625}	$&$0.020	$&$0.019	$&$ 0.001 $& ${1795}	$&${-0.625}	$&$0.057	$&$0.025	$&$ 0.003 $  \\ \hline

${1720}	$&${+0.625}	$&$0.012	$&$0.012	$&$ 0.001$ &${1815}	$&${+0.625}	$&$0.061	$&$0.051	$&$ 0.005 $ \\ 
${1720}	$&${+0.375}	$&$0.019	$&$0.017	$&$ 0.001 $& ${1815}	$&${+0.375}	$&$0.062	$&$0.052	$&$ 0.005 $ \\ 
${1720}	$&${+0.125}	$&$0.020	$&$0.021	$&$  0.001$ &${1815}	$&${+0.125}	$&$0.042	$&$0.037	$&$ 0.003 $ \\ 
${1720}	$&${-0.125}	$&$0.034	$&$0.020	$&$ 0.002 $&${1815}	$&${-0.125}	$&$0.057	$&$0.052	$&$ 0.004 $  \\ 
${1720}	$&${-0.375}	$&$0.031	$&$0.022	$&$  0.002$& ${1815}	$&${-0.375}	$&$0.094	$&$0.047	$&$ 0.007 $  \\ 
${1720}	$&${-0.625}	$&$0.031	$&$0.029	$&$ 0.002 $&${1815}	$&${-0.625}	$&$0.106	$&$0.053	$&$ 0.008 $  \\ \hline

${1750}	$&${+0.625}	$&$0.032	$&$0.027	$&$ 0.001 $ &${1835}	$&${+0.625}     $&$0.098	$&$0.066	$&$ 0.008 $ \\ 
${1750}	$&${+0.375}	$&$0.034	$&$0.031	$&$ 0.002$& ${1835}	$&${+0.375}	$&$0.047	$&$0.033	$&$ 0.004 $  \\ 
${1750}	$&${+0.125}	$&$0.031	$&$0.028	$&$ 0.001 $& ${1835}	$&${+0.125}	$&$0.143	$&$0.050	$&$ 0.012 $ \\ 
${1750}	$&${-0.125}	$&$0.027	$&$0.025	$&$  0.001$&${1835}	$&${-0.125}	$&$0.069	$&$0.054	$&$ 0.006 $  \\ 
${1750}	$&${-0.375}	$&$0.036	$&$0.027	$&$ 0.002$& ${1835}	$&${-0.375}	$&$0.076	$&$0.052	$&$ 0.007 $ \\ 
${1750}	$&${-0.625}	$&$0.052	$&$0.045	$&$ 0.002 $& ${1835}	$&${-0.625}	$&$0.066	$&$0.060	$&$ 0.006 $ \\ \hline
 	
${1775}	$&${+0.625}	$&$0.049	$&$0.035	$&$ 0.002$ &${1855}	$&${+0.625}	$&$0.136	$&$0.072	$&$ 0.012 $  \\ 
${1775}	$&${+0.375}	$&$0.086	$&$0.032	$&$ 0.005 $& ${1855}	$&${+0.375}	$&$0.120	$&$0.049	$&$ 0.011 $  \\
${1775}	$&${+0.125}	$&$0.064	$&$0.035	$&$ 0.003 $& ${1855}	$&${+0.125}	$&$0.118	$&$0.048	$&$ 0.010 $ \\ 
${1775}	$&${-0.125}	$&$0.045	$&$0.032	$&$ 0.002 $ &${1855}	$&${-0.125}	$&$0.120	$&$0.042	$&$ 0.010 $  \\ 
${1775}	$&${-0.375}	$&$0.089	$&$0.039	$&$ 0.005 $ &${1855}	$&${-0.375}	$&$0.140	$&$0.041	$&$ 0.012 $ \\ 
${1775}	$&${-0.625}	$&$0.082	$&$0.032	$&$ 0.004 $&${1855}	$&${-0.625}	$&$0.176	$&$0.048	$&$~ 0.015 $  

	\label{E}
\end{tabular}
\end{ruledtabular}
\end{table*}

\begin{table}[htb]

\caption{Integrated cross section for $\gamma n \rightarrow K^0 \Sigma^0$.}
\begin{ruledtabular}
\begin{tabular}{cccc}

$W$ & ${\sigma }$& stat.\ unc. & sys.\ unc.\\
(\rm MeV)   & $(\mu b)$& ($\mu$b) & ($\mu$b) \\
\hline

$1690$	&    ${0.111 }$&$ 0.038$&$ 0.005 $ \\  
$1720$	&    ${0.338 }$&$0.067 $&$  0.011 $\\  
$1750$	&    ${0.47 }$&$0.12$&$ 0.01 $ \\  
$1775$	&    ${0.98 }$&$ 0.16$&$ 0.03 $ \\  
$1795$	&    ${0.54 }$&$ 0.11$&$0.02  $ \\  
$1815$	&    ${0.90}$&$0.23$&$0.06 $ \\  
$1835$	&    ${1.01}$&$ 0.23$&$ 0.08$\\  
$1855$	&    ${1.71 }$&$0.22$&$~ 0.13 $

	\label{F}
\end{tabular}
\end{ruledtabular}
\end{table}


\bibliography{basename of .bib file}

\end{document}